\documentclass[12pt,epsfig]{article} 
\input{epsf.tex}
\usepackage{amssymb,epsfig}
\newcommand{\One}{1\kern-4.5pt1}
\newcommand{\lapprox}{\raisebox{-0.5ex}{$\ 
\stackrel{\textstyle<}{\textstyle\sim}\ $}}
\newcommand{\gapprox}{\raisebox{-0.5ex}{$\ 
\stackrel{\textstyle>}{\textstyle\sim}\ $}}
\begin{document}

\addtolength{\baselineskip}{0.20\baselineskip}

\hfill June 2007

\vspace{48pt}

\centerline{\Huge Can we study Quark Matter} 
\centerline{\Huge in the Quenched Approximation?}

\vspace{18pt}

\centerline{\bf Pietro Giudice$^a$ and Simon Hands$^b$} 

\vspace{15pt}

\centerline{$^a$ {\sl 
Dipartimento di Fisica Teorica, Universit\`a di Torino and INFN, Sezione di
Torino,}} 
\centerline{\sl via P. Giuria 1, I-10125 Torino, Italy.}
\smallskip
\centerline{$^b$ {\sl Department of Physics, Swansea University,}}
\centerline{\sl Singleton Park, Swansea SA2 8PP, U.K.}
\vspace{24pt}


\centerline{{\bf Abstract}}

\noindent
{\narrower
We study a quenched SU(2) lattice gauge theory 
in which, in an attempt to distinguish between timelike and spacelike 
gauge fields, the gauge ensemble $\{U_\mu\}$ is generated 
from a 3 dimensional gauge-Higgs model, the
timelike link variables being ``reconstructed'' from the Higgs fields. 
The resulting 
ensemble is used to study quenched quark propagation with non-zero chemical 
potential $\mu$; in particular, the quark density, chiral and superfluid
condensates, 
meson, baryon and gauge-fixed quark propagators are all studied as functions
of $\mu$. While it proves possible to
alter the strength of the inter-quark interaction by changing the parameters of
the dimensionally reduced model, there is no evidence for any region of
parameter space where quarks exhibit deconfined behaviour or thermodynamic
observables scale as if there were a Fermi surface.
}




\vfill
\newpage
\section{Introduction}
\label{sec:introduction}

Lattice QCD at non-zero quark chemical potential $\mu$ ought in principle to be
much more straightforward than the corresponding problem with non-zero 
temperature $T$.
The reason is that $\mu$ can be introduced via a local term in the
Lagrangian, whereas $T>0$ is imposed via a lattice of finite extent 
$L_ta=T^{-1}$ in
the temporal direction.
Simulations with varying $\mu$ can therefore be performed at
fixed lattice spacing $a$, so that the renormalisation factors needed for bulk
thermodynamic observables such as the Karsch 
coefficients required to extract the physical energy density need be calculated
only once, rather than for each value of $T$ studied. Moreover, the regime
$\mu/T\gg1$ relevant for the physics of the superdense matter found in neutron
star cores implies lattices with large $L_t$, so that
temporal correlators can be sampled efficiently without recourse to 
anisotropic lattices requiring careful callibration. This means that 
excitations above the ground state of the system, or more generally the nature
of the spectral density function, can be studied with relative ease. 
Indeed, this
programme has been successfully carried out in certain theories
such as the NJL and related models, where both 
measurement of the superfluid gap 
in the quasiparticle
spectrum at $k\approx\mu$~\cite{HW} 
and the identification of a phonon excitation with 
$\omega\propto k$ in the spin-1
meson channel~\cite{HS} have proved possible.

In practice, of course, 
the reason this happy state of affairs has not been exploited in
QCD is
the notorious {\em Sign Problem} associated with most Euclidean field theories
having a non-zero density of a conserved charge. In brief, for lattice QCD with
$N_f$ quark flavors described by a 
Lagrangian density of the form $\bar q Mq$, the
functional measure $\mbox{det}^{N_f}M(\mu)=\mbox{det}^{N_f}M^*(-\mu)$ 
implying that for
$\mu\not=0$ the action is complex, rendering Monte Carlo importance
sampling impracticable in the thermodynamic limit. Can one then at least 
perform lattice QCD simulations in the quenched $N_f\to0$ limit, i.e. study the
propagation of valence quarks with $\mu\not=0$ through a non-perturbative gluon
background?

In principle the information extracted from such an approach could be at best
qualitative, since (unlike the case of $T>0$) 
the gauge field ensemble $\{U_\mu\}$ can only 
respond to $\mu\not=0$ via virtual quark loops, so that in an orthodox 
quenched simulation the gluon background is that of the vacuum with zero baryon
charge density. 
Nonetheless, such information
might be valuable, for instance, in furnishing a non-perturbative definition of
the Fermi surface, whose existence is assumed in most phenomenological
treatments of dense matter. 

In the NJL studies referred to above, the Fermi surface 
appears as a minimum of the dispersion relation $\omega(k)$ in the
neighbourhood of the Fermi momentum $k=k_F$: for weakly-interacting massless
quarks we expect $k_F\approx\mu$. Even free quark
propagation at $\mu\not=0$ reveals the presence of a Fermi surface 
and so is not entirely trivial. 
For QCD however,
the notion of a distinguished quark momentum is not 
gauge invariant, so that the identification of regions of 
phase space occupied by
low-energy excitations may be altogether more subtle in a gauge invariant
formalism.

However, as plausible as this sounds, the quenched approach has conceptual 
difficulties at $\mu\not=0$. In the context of a random matrix theory,
Stephanov~\cite{Misha} showed that the quenched theory should be thought of 
as the $N_f\to0$ limit of a QCD-like theory with not just $N_f$ flavors of 
quark $q\in{\bf3}$ of the SU(3) gauge group 
but also with $N_f$ flavors of conjugate quark $q^c\in\bar{\bf3}$. As has been
known for many years~\cite{Gocksch}, the extra particle content results in
gauge invariant $qq^c$ bound states in the spectrum which
in a strongly interacting theory can result in baryons degenerate with 
light mesons. At $\mu=0$ these extra states are usually regarded
as extra pions due to the fermion species doubling resulting from the
use of the manifestly real positive measure $\mbox{det}MM^\dagger$ required by
practical fermion algorithms. The introduction of $\mu\not=0$ distinguishes
baryons from mesons. A simple argument, which assumes
that binding energy is a relatively small component of the energy density of
bulk nuclear matter, predicts that for
$\mu/T\gg1$ there should be an {\em onset transition} 
from the vacuum to a ground state with
quark number density $n_q>0$ once $\mu$ equals the mass of the lightest baryon
divided by the number of quark constituents. For QCD this scale
is $m_N/3$, where $m_N$ is the nucleon mass; for a theory with conjugate quarks
the onset scale is $m_\pi/2$. Only cancellations among configurations 
due to a 
complex-valued measure $\mbox{det}M$ 
can ensure that the fake signal for $n_q>0$
vanishes in the range $m_\pi/2<\mu<m_N/3$~\cite{TCQCD}. 
A recent analytic demonstration 
has been given, once again in the context of a random matrix model, in 
\cite{RMT}.

It is clear from these considerations that quenched calculations at $\mu\not=0$
can only be useful if the distribution of gluon fields is modified in some way
to reflect high baryon density.  For instance,
a minimum requirement is that the timelike
links $U_0$ are no longer drawn from the same sampling
distribution as spacelike 
links $U_i$, reflecting the breakdown of Lorentz invariance due to the preferred
rest frame of the background quark distribution.
If the gluons were modified in 
some way so that color confinement no longer holds, then the role of $qq^c$
excitations may not be so important in determining the ground state in the 
quark sector, 
and it is at least conceivable that valence
quark propagation in such a background may qualitatively resemble that
of the deconfined regime of the phase diagram at $\mu/T\gg1$ corresponding to
quark matter. The focus of such a study would thus be ``cold quarks in hot
glue''. While there are several lines one could take, the approach we shall
explore in this study is to start with the 3$d$ configurations characteristic 
of the deconfined phase found at $T>T_c$, $\mu/T\ll1$
produced by the approach to
hot gauge theory known as {\em Dimensional Reduction\/} (to be reviewed below in
Secs.~\ref{sec:4D} and \ref{sec:parameters}). 
In this approach all 
non-static modes of the gauge theory
(ie. those with non-zero Matsubara frequency) are integrated
out leaving a 3$d$ gauge-Higgs model describing the non-perturbative behaviour
of the remaining static modes. For sufficiently large $T/T_c$ the model
coefficients
are perturbatively calculable functions of $T$,
$\mu$ and $N_f$, and the resulting effective theory can be used to make
quantitative predictions in the quark gluon plasma phase.
Our goals are less ambitious and more speculative; we will use the Higgs fields
of the 3$d$ theory to ``reconstruct'' the timelike gauge fields $U_0$ and hence
generate (3+1)$d$ configurations of static gauge fields suitable for the study 
of valence quark propagation with $\mu\not=0$. At this stage we are simply
trying to generate a non-confining gluon background, and in no sense claim 
to be developing a high density effective theory.

With gauge group SU(3) and $\mu\not=0$, 
the dimensional reduction (DR) machinery yields a cubic
higgs self-interaction with imaginary coefficient proportional to $N_f$
~\cite{HLP}, 
which is how the 
3$d$ effective theory inherits the sign problem from (3+1)$d$ QCD. Whilst there is
no reason to suppose the sampling of the model's configuration space would be
any less problematic than that of the full theory, at least the complex phase of
each configuration is now expressed in terms of a local term in the
action, making its
calculation cheap. Moreover, since the coefficient of the complex term can 
be
made arbitrarily small {\em ad hoc\/}, 
the effect of gradually introducing the sign problem,
and the interplay of the complex phase with physical observables, could be
explored. In this paper, however, we focus on the technically simpler case of
gauge group SU(2). In this case the action is real for even $N_f$ (and the cubic
DR higgs self-interaction vanishes), permitting orthodox lattice simulations 
with standard algorithms, e.g.~\cite{HKLM}.

QCD with gauge group SU(2), 
often referred to as Two Color QCD or QC$_2$D, has been
studied at $\mu\not=0$ with lattice simulations
by several groups using a variety of formulations and 
algorithms~\cite{TCQCD,HKLM,TCQCD2,TCQCD3,HKS,ADL}. 
Since $q$ and $\bar q$ fall in equivalent
representations of the gauge group, so that in effect $q$ and $q^c$ are
identical, hadron multiplets contain both $q\bar q$
mesons and $qq$, $\bar q\bar q$ baryons, which are degenerate at $\mu=0$. In the
chiral limit the lightest hadrons are Goldstone bosons, and can be analysed
using chiral perturbation theory ($\chi$PT) \cite{KSTVZ}. At leading order
for $\mu\gg T$ a second order 
onset transition from vacuum to matter consisting of tightly bound diquark
scalar bosons is predicted at exactly $\mu_o=m_\pi/2$. At the same point
the chiral condensate
$\langle\bar qq\rangle$ starts to fall below its vacuum value, and a
non-vanishing diquark condensate $\langle qq\rangle$ develops, such that
$\langle\bar qq\rangle^2+\langle qq\rangle^2$ remains constant. The diquark
condensate spontaneously breaks U(1) baryon number symmetry, so
the resulting ground state is superfluid. In the limit $\mu\to\mu_{o+}$ the
matter in the ground state becomes arbitrarily dilute, weakly-interacting, and
non-relativistic, 
and is a textbook
example of Bose-Einstein condensation. This scenario was subsequently confirmed
by simulations with staggered lattice fermions~\cite{TCQCD,TCQCD2}.
More recent simulations have found evidence for a second transition at larger
$\mu$ to a deconfined phase, as evidenced by a non-vanishing Polyakov
loop~\cite{HKS} and by a fall in the topological susceptibility~\cite{ADL}.
In this regime thermodynamic quantities scale according to the expectations of 
free field theory (also referred to as ``Stefan-Boltzmann'' (SB) scaling), 
namely $n_q\propto\mu^3$, and energy density 
$\varepsilon\propto\mu^4$~\cite{HKS}.

In this paper we build on our existing experience by exploring $\mu\not=0$ in
quenched QC$_2$D. In Sec.~\ref{sec:formul} below we specify our procedure for 
generating (3+1)$d$ quenched SU(2) gauge configurations starting from a 3$d$
gauge-Higgs model, and review standard quark observables once $\mu\not=0$.
Our results follow in Sec.~\ref{sec:results}. First we explore the quark density
$n_q$ and the chiral $\langle\bar qq\rangle$ and superfluid  $\langle qq\rangle$
condensates to see how the
equation of state responds to attempts to render the (3+1)$d$ theory
``non-confining'', and whether the behaviour predicted by $\chi$PT can be
supplanted by the SB scaling expected of weakly-interacting
degenerate quarks. Next we turn to spectroscopy, calculating
both ``normal'' and ``anomalous'' components of the quark 
propagator, taking advantage of the
enormous gain in statistical accuracy 
offered by the quenched approach. First we study 
the bound state spectrum in both meson and diquark sectors, finding evidence for
significant mixing between the sectors at large enough $\mu$. Next, for the
first time in a gauge theory, we present results for the quark
propagator at $\mu\not=0$ obtained following gauge-fixing. 
A range of spatial momenta are sampled 
in an attempt to map out the quasiquark dispersion relation. 
While we find significant qualitative differences between strongly and
weakly-interacting quarks, no signal for a Fermi surface has emerged.
Our conclusions follow in Sec.~\ref{sec:conclusions}.

\section{Formulation}
\label{sec:formul}

\subsection{Reconstructing the Fourth Dimension}
\label{sec:4D}
The quenched action we start from is the 3$d$ SU(2) 
gauge -- adjoint Higgs model given by
Eqn. (4) of Ref.\cite{HP1}:
\begin{eqnarray}
S_{3d}&=&\beta\sum_{x,i>j}\left(1-{1\over2}\mbox{tr}U_{x,ij}\right)+
2\sum_x\mbox{tr}(\varphi_x\varphi_x)\nonumber\\
&-&2\kappa\sum_{x,i}\mbox{tr}(\varphi_x
U_{x,i}\varphi_{x+\hat\imath}U_{x,i}^\dagger)+
\lambda\sum_x(2\mbox{tr}(\varphi_x\varphi_x)-1)^2,\label{eq:S3d}
\end{eqnarray}
where $\varphi\equiv{1\over2}\varphi_a\tau_a$ is a traceless hermitian
$2\times2$ matrix representing the adjoint Higgs field. 
As is usual in a gauge-Higgs model, increasing $\kappa$ at sufficiently large
$\beta$ takes one from a ``confinement'' phase with small
$\langle\mbox{tr}\varphi\varphi\rangle$ to a ``Higgs'' phase with large
$\langle\mbox{tr}\varphi\varphi\rangle$. 

The action (\ref{eq:S3d}) is derived by dimensional reduction from a 4$d$
pure gauge SU(2) theory with non-zero temperature $T$. In continuum notation
the original Lagrangian
density ${\cal L}_4\sim(\partial_\mu A_\nu)^2$
in terms of a gluon field with 
canonical mass dimension $[A]=1$. The dimensionally reduced Lagrangian is
obtained by integrating over Euclidean time and discarding all non-static modes,
ie. setting $\partial_0=0$:
\begin{equation}
{\cal L}_3=\int_0^{1\over T}dx_0{\cal L}_4\sim{1\over T}(D_i A_j)^2+
{1\over T}(D_i A_0)^2\equiv(D_i B_j)^2+(D_i\Phi)^2,
\end{equation}
with the new fields defined by $B_i=T^{-{1\over2}}A_i$,
$\Phi=T^{-{1\over2}}A_0$, $[B]=[\Phi]={1\over2}$.
When transcribing to the 3$d$ lattice action (\ref{eq:S3d}) we use
\begin{equation}
U_i=\exp(igaA_i)=\exp(igaT^{1\over2}B_i);\;\;\;
\varphi=\sqrt{a\over\kappa}\Phi,
\label{eq:rescaling}
\end{equation}
with $[U]=[\varphi]=0$. The 3$d$ lattice parameters $\beta$, $\kappa$ and
$\lambda$ are all dimensionless, with $\beta$ given in terms of the 
Yang-Mills coupling by
\begin{equation}
\beta={4\over{ag_3^2}}\equiv{4\over{ag^2T}}={4N_\tau\over g^2},
\label{eq:couplings}
\end{equation}
where it is helpful to distinguish between the dimensionful coupling $g_3$ of
the 3$d$ theory and the dimensionless coupling $g$ of the parent hot 4$d$
theory, related via $g_3^2=g^2T$; in the final equality we have assumed 
that the 4$d$ theory is formulated 
on a lattice with $N_\tau$ time spacings.

In the DR 
approach to the effective description of hot field
theory, the parameters $\beta$, $\kappa$ and $\lambda$ are perturbatively
calculable functions of $g$, $T$ and quark chemical potential $\mu$, 
as outlined in \cite{HLP}. The
attractiveness of this approach is that effects of, 
say $N_f$ flavors of massless 
quark or a small quark chemical potential $\mu$ can be incorporated in the
coefficient calculation, the result always being a $3d$ bosonic  model
which is relatively cheap to simulate. The DR provides a good effective
description when the non-static modes decouple; the authors of \cite{HLP,HP1}
claim that this is valid for $T\gapprox2T_c$, $\mu\lapprox4T$.
The parameter choice employed in this study is discussed further below in
Sec.~\ref{sec:parameters}.

As discussed in Sec.~\ref{sec:introduction},
the quenched approximation has long been considered to be inapplicable to
QCD with $\mu\not=0$, because it includes unphysical light baryons
formed as bound states of $q\in{\bf3}$ and
$q^c\in\bf{\bar3}$ \cite{Misha,Gocksch}. The reason these $qq^c$
states are light and hence 
distort the physics is because at $T=0$ quenched QCD configurations are in a
confining phase, implying spontaneous chiral symmetry breaking. The lightest
$qq^c$ state is degenerate with the Goldstone pion.
Our goal is to study 
quark propagation 
through a 
{\em non-confining\/} quenched gluon background with $\mu\not=0$. Since
chemical potential couples to quarks via the timelike component of the 
current $\mu\bar\psi\gamma_0\psi$, this is an
inherently four-dimensional problem. In order to 
generate such a background we take a 3$d$ configuration generated
by the DR simulation,
motivated by the fact that it describes deconfining physics, and
``reconstruct'' the gauge field in the timelike direction following
(\ref{eq:rescaling}) via the prescription (recall 
$aA_0=aT^{1\over2}\Phi=(a\kappa
T)^{1\over2}\varphi$):
\begin{equation}
U_0=
\exp\left(ig\sqrt{\kappa\over
N_\tau}\varphi\right)
=\cos\left(\tilde g\sqrt{\varphi_a\varphi_a}\right)
+i{{\tau_a\varphi_a}\over\sqrt{\varphi_a\varphi_a}}
\sin\left(\tilde g\sqrt{\varphi_a\varphi_a}\right),
\label{eq:extend}
\end{equation}
with $\tilde g=\sqrt{\kappa\over\beta}$.
Spatial link variables $U_i$ are taken to be time independent and identical 
to their 3$d$ counterparts.
The resulting model differs from DR in that it has a non-trivial electrostatic
potential (ie. a spatially-varying $U_0$ field), 
but more importantly differs from real physics with 
$\mu\gg T$ in that there are no 
non-static (ie. $\partial_0\not=0$) modes\footnote{Even the static modes in
dense matter with a sharp Fermi surface may exhibit oscillatory behaviour, 
known as
Friedel oscillations~\cite{friedel}, which are absent from the DR approach.}.
It is {\em not} 
derivable from QCD in 
any systematic way.

It is legitimate to ask whether excluding non-static modes from consideration is
too severe an approximation to be physically reasonable. It is possible to
examine this issue for degenerate quark matter interacting weakly via gluon
exchange ~\cite{KogSteph}. The loop integral in the self-consistent equation
for the color-superconducting gap $\Delta$
is dominated by small angle scattering between
quarks at antipodal points of the Fermi sphere, diverging as $\ln\theta$ where
$\theta\sim q_0/\mu$, $q_0$ being the timelike momentum of the exchanged
gluon. The divergence must be cut off by some physical screening mechanism: for
electric gluons this is Debye screening, resulting in a $\theta_{min}\sim g$
below which the interaction is effectively point-like; for magnetic gluons the
mechanism is Landau damping resulting in $\theta_{min}\sim
g^{2\over3}q_0^{1\over3}/\mu^{1\over3}$. In the static limit $q_0\to0$ magnetic
gluons are thus unscreened in perturbation theory, and the resulting gap
equation yields $\Delta\sim\mu\exp(-3\pi^2/\sqrt{2}g)$ \cite{Son}.
The effect of neglecting non-static modes such as electric gluons can be shown
to affect the pre-exponential factor, but not the scaling of the gap with the
coupling strength. Neglect of non-static modes thus seems defensible in a
weak-coupling approach to quark matter; 
beyond perturbation theory we have little to guide us.

A conceptual point worth stressing is that due to the different ways spacelike
and timelike links are treated, there is no simple relation between spatial
and temporal lattice spacings. In principle one
could determine the relation empirically 
by comparison of correlators in different
directions, and $a_s/a_t$ callibrated by matching, say, to the ratio of the
physical (and $T,\mu$-dependent) Debye screening mass to a hadron mass measured
at $\mu=0$.
We make no attempt to follow this prescription in this exploratory study: all
masses and energy scales are compared with a reference scale chosen to be the
pion mass at $\mu=0$, 
calculated in dimensionless units as $m_\pi a_t$. This prevents us
for the time being from a direct comparison with condensates such as the 
physically observable quark density, expressed in dimensionless units as
$n_q a_s^3$. Note, however, that the dimensionless ratio $\mu/T$ can be
expressed without ambiguity using lattice variables as $N_\tau\tilde\mu$, where 
$N_\tau$ is the temporal extent of the lattice and
$\tilde\mu=\mu a_t$ 
is the dimensionless lattice chemical potential introduced in
Eqn.~(\ref{eq:fermaction}) below.

To illustrate both the benefits and potential pitfalls of this approach
it is helpful to consider rectangular Wilson loops. The
spatial Wilson loop $W(r_1,r_2)$ clearly inherits the behaviour of the 3$d$
theory; it is known to decay with an area law \cite{Wilson}. Confinement in the
(3+1)$d$ theory, however, is governed by the temporal loop $W(r,t)$ given by
\begin{equation}
W(r,t)={1\over N_c}\mbox{tr}P(\vec 0,0;\vec r,0)
U_0^t(\vec r)P^\dagger(\vec 0,t;\vec
r,t)(U_0^\dagger(\vec 0))^t.
\end{equation}
Here $P$ is the path-ordered product of spacelike links connecting $\vec 0$ to
$\vec r$; because the configuration is static these can be written
$P(\vec 0,0;\vec r,0)=P(\vec 0,t;\vec r,t)\equiv P$. Using
(\ref{eq:couplings},\ref{eq:extend}) we deduce
\begin{equation}
W(r,t)={1\over N_c}\mbox{tr}Pe^{i2\tilde g t\varphi_0}P^\dagger
e^{-i2\tilde g
t\varphi_r}.
\end{equation}
Now expand the exponentials as power series in $\tilde g$, noting that
$(2\varphi)^2=\varphi_a\varphi_a\One\equiv\vert\varphi\vert^2\One$, and
$\mbox{tr}\varphi^n=0$ for odd $n$, implying that odd terms vanish. 
The lowest non-trivial term is $O(\tilde
g^2)$:
\begin{equation}
-{{\tilde g^2t^2}\over 2!}
\biggl(\vert\varphi_0\vert^2+\vert\varphi_r\vert^2\biggr)
+{{4\tilde g^2t^2}\over N_c}\mbox{tr}(P\varphi_0P^\dagger\varphi_r).
\label{eq:Og2}
\end{equation}
To proceed we make some simplifying approximations. For any 
$\varphi\not=0$, $r>0$, the second term of (\ref{eq:Og2}) is proportional to
$\varphi_0^a\varphi_r^bP_{\rm adj}^{ab}$; 
in the strong coupling limit $\beta\to0$, strong 
link field fluctuations should make its expectation value
very small, and hence we neglect all terms of this form. Secondly,
when integrating over fluctuations of the $\varphi$ we assume translational
invariance and neglect all non-Gaussian
fluctuations (which should be valid as $\lambda\to0$), so that
$\langle\vert\varphi_0\vert^{2p}\vert\varphi_r\vert^{2q}\rangle\approx\langle
\vert\varphi\vert^{2(p+q)}\rangle$. This also requires
physical excitations of the $3d$ theory to be massive so that
$\langle\varphi_0\varphi_r\rangle$ can be neglected.
It is then straightforward to show that all
$r$-dependence drops out and the 
expectation value of the 
$O(\tilde g^{2n})$ term is $(-1)^n(2\tilde
gt\langle\vert\varphi\vert\rangle)^{2n}/(2\times(2n)!)$. Hence
\begin{equation}
\langle W(r,t)\rangle\approx \cos^2(\tilde gt\langle\vert\varphi\vert\rangle).
\end{equation}
As required, the temporal Wilson loop is positive, well-behaved for 
$\tilde gt\vert\varphi\vert\ll1$, and clearly does not obey an
area law. The reconstructed (3+1)$d$
theory is thus non-confining. 
Note that the neglected $P\varphi P^\dagger\varphi$ 
terms always contribute to the series coefficients with the opposite sign (due
to an extra factor of $i^2$), so that their inclusion would in effect
make $\langle W\rangle$ decay less rapidly.
Viewed as a function of Euclidean time $\langle W\rangle$ 
decays with a negative
curvature; this is inconsistent with a transfer matrix with positive definite
spectrum, which implies decays of the form $\sum_ic_ie^{-E_it}$. 
Unsurprisingly, therefore, the (3+1)$d$ model violates unitarity.

\subsection{Parameter Choice for the Dimensionally Reduced Model}
\label{sec:parameters}

The DR procedure for SU(N) gauge theory with $N_f$ flavours of 
fermions has been studied to two loops in \cite{HP1}.
The effective theory obtained is a three dimensional SU(N) adjoint Higgs 
theory, with the scalars corresponding to the electric
gauge potential $A_0$ in the unreduced theory; the case with chemical
potential $\mu\not=0$ is dealt with for example in 
Ref.~\cite{HLP}. 

The continuum action of the SU(2) adjoint Higgs model is given 
by~\cite{HP1}
\begin{equation}
S=\int d^3x \left\{\frac{1}{2}\mbox{tr}\left(F_{ij}F_{ij}\right) 
+\mbox{tr}\left( D_i\Phi D_j\Phi \right)
+m_3^2 \mbox{tr}\left( \Phi\Phi \right)
+\lambda_3 \left[  \mbox{tr}\left( \Phi\Phi \right) \right]^2
  \right\},
\label{action_su2_continuum}
\end{equation}
where $F_{ij}=\partial_i A_j-\partial_j A_i+ ig_3[A_i,A_j]$ and 
$D_i\Phi=\partial_i\Phi+ig_3 [A_i,\Phi]$. 
The physical properties of the theory are fixed by the two dimensionless ratios
\begin{equation}
x=\frac{\lambda_3}{g_3^2}, \qquad y=\frac{m_3^2}{g_3^4}.
\end{equation}
In the framework of DR these parameters are
completely determined by $g^2$ and $T$ of the original
four dimensional gauge theory.
Since $g^2$ is a running coupling its value is fixed by the 
renormalisation scale $\Lambda_{\overline{\rm MS}}$.
Choosing the renormalisation scale as in~\cite{kalarusha97} and
expressing $\Lambda_{\overline{\rm MS}}$ through the critical temperature 
$T_c=1.23(11) \Lambda_{\overline{\rm MS}}$
as measured on the lattice~\cite{fiheka93}, 
it is possible to show~\cite{kalarusha97} that for $N_f=0$
\begin{equation} 
g_3^2=\frac{10.7668}{\ln(8.3T/T_c)} T, 
\quad x=\frac{0.3636}{\ln(6.6T/T_c)}, 
\quad y(x)=\frac{2}{9\pi^2 x}+\frac{1}{4\pi^2} + {\cal O}(x).
\label{eq_4d_3d}
\end{equation}
With these equations, specifying $T/T_c$ completely 
fixes the parameters $x,y$ of the reduced model.
Note that the values of $x$ and $y$ corresponding to $T_c$ are about
$x_c\approx0.19$ and $y_c\approx0.14$.
The model exhibits two phases: a symmetric phase with full SU(2)-type
confinement and a Higgs phase with residual U(1)-type 
confinement~\cite{nadkarni}.

Simulations of the three dimensional SU(2) adjoint Higgs model 
showed that this theory has a phase diagram with continuously connected
Higgs and confinement phases, which are partially separated by a line of 
first order phase transitions~\cite{kalarusha97,haphtest97}.
The discretized form of the action (\ref{action_su2_continuum}) in terms of a
rescaled lattice field $\varphi$ is simply our action (\ref{eq:S3d}).
The $3d$ lattice parameters $\beta, \kappa$ and $\lambda$ are all 
dimensionless.
The parameters of the continuum and lattice theory are related
up to two loops
by a set of equations~\cite{laine95},
\begin{eqnarray}
x&=& \frac{\beta \lambda}{\kappa^2},  \qquad 
\beta=\frac{4}{ag_3^2},  \nonumber\\
y&=& \frac{\beta^2}{8}
                \left(\frac{1}{\kappa}-3
                -\frac{2x\kappa}{\beta}\right) 
                +\frac{\Sigma\beta}{4\pi}
                     \left(1+\frac{5}{4}x\right) \nonumber\\
                &+&\frac{1}{16\pi^{2}}
                  \left[(20x-10x^2)
                  \left(\ln\left(\frac{3\beta}{2}\right)+0.09\right)
                  +8.7+11.6x\right],
\label{eq_cont_latt}
\end{eqnarray}
where $\Sigma = 3.17591$.  
Due to the theory's superrenormalisability, these perturbative 
relations are exact in the continuum limit and, based on experience,
are accurate for all  $\beta > 6$~\cite{HP1}. 
\begin{table}[t]
\begin{center}
\begin{tabular}{|c|c|c|c|c|}
\hline
$T/T_c$ & $x$ & $y$ & $\kappa$ & $\lambda$  \\
\hline
$1$    & 0.193 & 0.142 & 0.3637812 & 0.0028379  \\
$2$    & 0.141 & 0.185 & 0.3620027 & 0.0020531  \\
$3$    & 0.122 & 0.210 & 0.3612335 & 0.0017689  \\
$5$    & 0.104 & 0.242 & 0.3603992 & 0.0015009  \\
$10$   & 0.087 & 0.285 & 0.3594500 & 0.0012490  \\
$100$  & 0.056 & 0.427 & 0.3569257 & 0.0007927  \\
\hline
\end{tabular}
\caption{The lattice parameters for $\beta=9.0$. \label{tab_params} }
\end{center}
\end{table}
In Tab.~\ref{tab_params} we report some sample lattice parameter sets 
obtained by calculating $x$ and $y$ according to Eqs.~(\ref{eq_4d_3d}) and 
fixing the 
ratio $T/T_c$; then we have calculated $\kappa$ and $\lambda$ using
Eqs.~(\ref{eq_cont_latt}).

For gauge group SU(3) the DR theory is very similar to 
Eq.~(\ref{action_su2_continuum})~\cite{HLP}:
\begin{equation}
  S = \int d^{3}x \left\{ \frac{1}{2} \mbox{tr} F_{ij}^2
  +\mbox{tr} [D_{i},A_0]^2 +m_3^{2}  \mbox{tr} A_0^2 
  +\lambda_3( \mbox{tr} A_0^2)^{2} \right\},
\label{su3_action_cont}
\end{equation}
where $F_{ij}=\partial_{i}A_{j}-\partial_{j}A_{i}
 +ig[A_i,A_j]$,
$D_i = \partial_i + ig_3 A_i$, 
$F_{ij},A_i$, and $A_0$ are all traceless $3\times 3$
Hermitian matrices ($A_0=A_0^{a}T_{a}$, etc), and $g_3^2$ and $\lambda_3$
are the gauge and scalar coupling constants, with mass dimension one.
The physical properties of the effective theory, also in this case, 
are determined by the dimensionless ratios 
\begin{equation}
x=\frac{\lambda_3}{g_3^2}, \quad y=\frac{m_3^2(\bar\mu_3=g_3^2)}{g_3^4},
\label{param_su3}
\end{equation}
where $\bar\mu_3$ is the $\overline{MS}$ dimensional regularization
scale in $3d$. 

Here we note that application of DR to QCD with $N_f$ quark
flavors and
non-zero chemical potential $\mu$ results in a complex term
$ig^3\mu{N_f\over{3\pi^2}}\mbox{tr}A_0^3$~\cite{HLP}, 
so that the DR theory inherits the 
complex action of the parent theory (for gauge group SU(2)
$\mbox{tr}A_0^3\equiv0$ and the action is real). In principle a term of this
form could be incorporated in our approach, the advantage being that a ``Sign
Problem'' could be introduced with arbitrarily small coefficient, and the
response of observables to small phase fluctuations of the measure assessed. 
This gradualist approach is of course impossible in full QCD simulations.

In applications of the DR action (\ref{eq:S3d}) to hot gauge theory, 
the parameter $\beta$ is used to control the physical lattice spacing in units
of $T_c^{-1}$,
and $\kappa$ and $\lambda$ are tuned according to relations
(\ref{eq_4d_3d},
\ref{eq_cont_latt}) to specify $T$.
A complication
\cite{HP1} is that for the $\{\beta,\kappa,\lambda\}$ parameter set 
appropriate 
for hot QCD the 3$d$ model has its true ground state in the ``Higgs'' phase,
whereas the perturbative continuation to QCD 
(\ref{eq_4d_3d}) requires it to be in the ``confining'' phase\footnote{In this 
context the terms ``confining'' and ``Higgs'' refer to 3$d$
dynamics, corresponding to the behaviour of {\em spatial} Wilson loops in a 
(3+1)$d$ model. 
They have no bearing on the behaviour either of the hot gauge theory
before DR or the reconstructed 4$d$ model studied here.}.
Fortunately the ``confining'' 
phase appears to be metastable, so that simulations
started off with small $\langle\mbox{tr}\varphi\varphi\rangle$ can be used to 
yield physically relevant results. 
Once the model is extended to 4$d$ using (\ref{eq:extend}) the physical
meaning of the lattice parameters is no longer clear. In this exploratory study
we hold $\beta,\lambda$ fixed and, with the exception of
Figs.~\ref{fig:nqvslambda_mukap} 
and \ref{fig:nqvsmu_lamkap} below, 
restrict our attention to two values of $\kappa$.  

\subsection{Introducing Quarks with $\mu\not=0$}

Henceforth we use 4$d$ configurations $\{U_\mu\}$ generated 
as outlined above as input in quenched studies using 
staggered fermions in the fundamental representation of SU(2), having action
$S_1$+$S_2$ with
\begin{equation}
S_1={1\over2}\sum_{x\nu}\bar\chi_x
[e^{\tilde\mu\delta_{\nu0}}U_{x,\nu}\chi_{x+\hat\nu}
-e^{-\tilde\mu\delta_{\nu0}}U^\dagger_{x-\hat\nu,\nu}\chi_{x-\hat\nu}]+m\sum_x
\bar\chi_x\chi_x\equiv\bar\chi M\chi
\label{eq:fermaction}
\end{equation}
and
\begin{equation}
S_2=\sum_x
{j\over2}\chi^{tr}_x\tau_2\chi_x+
{{\bar\jmath}\over2}\bar\chi_x\tau_2\bar\chi_x^{tr}.
\label{eq:dsource}
\end{equation}
As explained above, the lattice parameter $\tilde\mu$ is related to the
physical quark chemical potential via $\tilde\mu=\mu a_t$. Henceforth we will
ignore the distinction between $a_s,\,a_t$ when quoting values for parameters
such as quark mass $m$ and diquark source $j$; in consequence 
no attempt will be made to compare these quantities with a physical scale.

The simplest observables to discuss are the chiral condensate and quark
density:
\begin{equation}
\langle\bar\chi\chi\rangle={1\over 2V}\Bigl\langle\mbox{tr}
{{\partial {\cal G}^{-1} }\over{\partial
m}}{\cal G}\Bigr\rangle;\;\;\;
n_q={1\over 4V}\Bigl\langle\mbox{tr}
{{\partial {\cal G}^{-1} }\over{\partial\tilde\mu}}{\cal G}\Bigr\rangle,
\end{equation}
where $\langle\ldots\rangle$ denotes a quenched average using the action 
(\ref{eq:S3d}), and ${\cal G}$  is the Gor'kov propagator defined below in 
(\ref{eq:Gor'kov}).
The normalisation, somewhat arbitrary in a quenched simulation, is
chosen so that in the limit $\mu\to\infty$ $n_q$ saturates at a value of
two per lattice site.
The term $S_2$ comprising gauge invariant diquark source 
terms\footnote{Note that a gauge invariant diquark
source does not exist for gauge group SU(3). Physically, this means diquark
condensation forces {\em superfluidity} for SU(2), but {\em superconductivity}
for SU(3).
Technically, it means diquark condensation is more difficult to address by
lattice methods for SU(3).} 
is introduced
in order to discuss the possibility of
diquark condensation. The Pauli matrices
$\tau_2$ act on SU(2) color indices.
While $S_1$ is invariant under the global U(1)
rotation $\chi\mapsto e^{i\alpha}\chi$, $\bar\chi\mapsto\bar\chi e^{-i\alpha}$,
$S_2$ is not.
Defining source strengths $j_\pm=j\pm\bar\jmath$, 
we obtain diquark condensates
\begin{equation}
\langle qq_\pm\rangle={1\over 2V}\Bigl\langle\mbox{tr}
{{\partial {\cal G}^{-1} }\over{\partial
j_\pm}}{\cal G}\Bigr\rangle.
\label{eq:qq}
\end{equation}
If we choose sources such that $j_+=j_+^*=j$, $j_-=0$, then the condensate
forms in the ``+'' channel, and a massless Goldstone pole develops 
in the ``--'' channel
as $j\to 0$ as confirmed by the Ward Identity
\begin{equation}
\sum_x\langle qq_-(0)qq_-(x)\rangle={{\langle qq_+\rangle}\over j_+}.
\end{equation}
The numerical implementation of all these observables is identical to that for
the NJL model described in \cite{HW}.

\section{Numerical Results}
\label{sec:results}

We have chosen $\beta=9.0$ sufficiently close to the continuum limit for the 
DR formalism to be trustworthy, and
start with a point with $T=2T_c$ according to
(\ref{eq_4d_3d}); from Table~\ref{tab_params} this corresponds to 
$\kappa=0.3620027$, $\lambda=0.0020531$.
\begin{figure}
\begin{center}
\epsfig{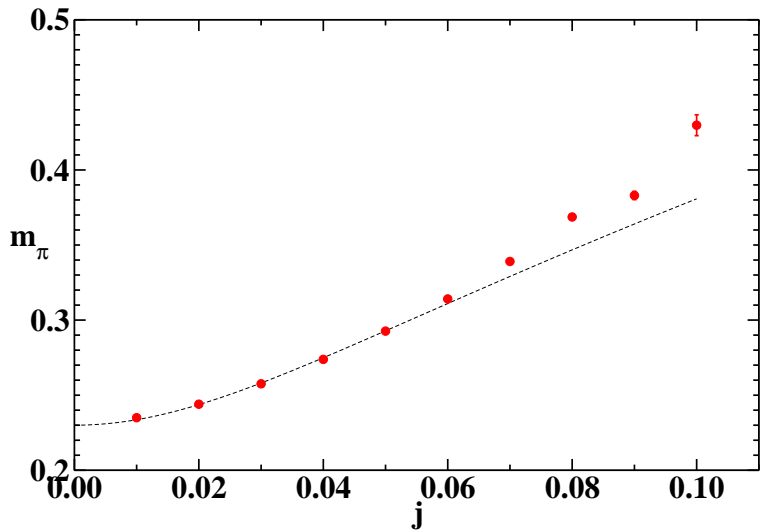}
\smallskip
\caption{$m_\pi$ vs. $j$ from data taken on $8^3\times128$ at $ma=0.05$, 
$\tilde\mu=0$
for $\beta=9.0$, $\kappa=0.362007$, $\lambda=0.0020531$.}
\label{fig:pionmass}
\end{center}
\end{figure}
For this parameter set, we have explored system volumes $8^3\times L_t$, 
with $L_t=16$, 32,  64 and 128. 
For the most part in this section 
we present data taken with $L_t=32$. However, it is important to make a precise 
determination of the pion mass at $\mu=0$.
Fig.~\ref{fig:pionmass} shows data for the pion mass $m_\pi$ as a function of
diquark source $j$ taken at $\mu=0$. The data are extrapolated to $j=0$ using a
$\chi$PT-inspired form \cite{KSTVZ} $m_\pi(j)=m_\pi(0)(1+bj^2)^{1\over4}$
(the bare quark mass $ma=0.05$, so that the pion remains massive as $j\to0$).
We conclude $m_\pi a_t(j=0)=0.230(3)$, 
consistent with a measurement made exactly at
$\mu=j=0$: $m_\pi a_t=0.2321(1)$. Since this scale is not too dissimilar to
$L_s^{-1}$, we have repeated the measurement on $16^3\times64$, where we find
$m_\pi a_t=0.2368(3)$. The systematic error due to finite volume is significant,
but small enough at 2\% to be acceptable for this exploratory study.

\subsection{Equation of State}

\begin{figure}
\begin{center}
\epsfig{file=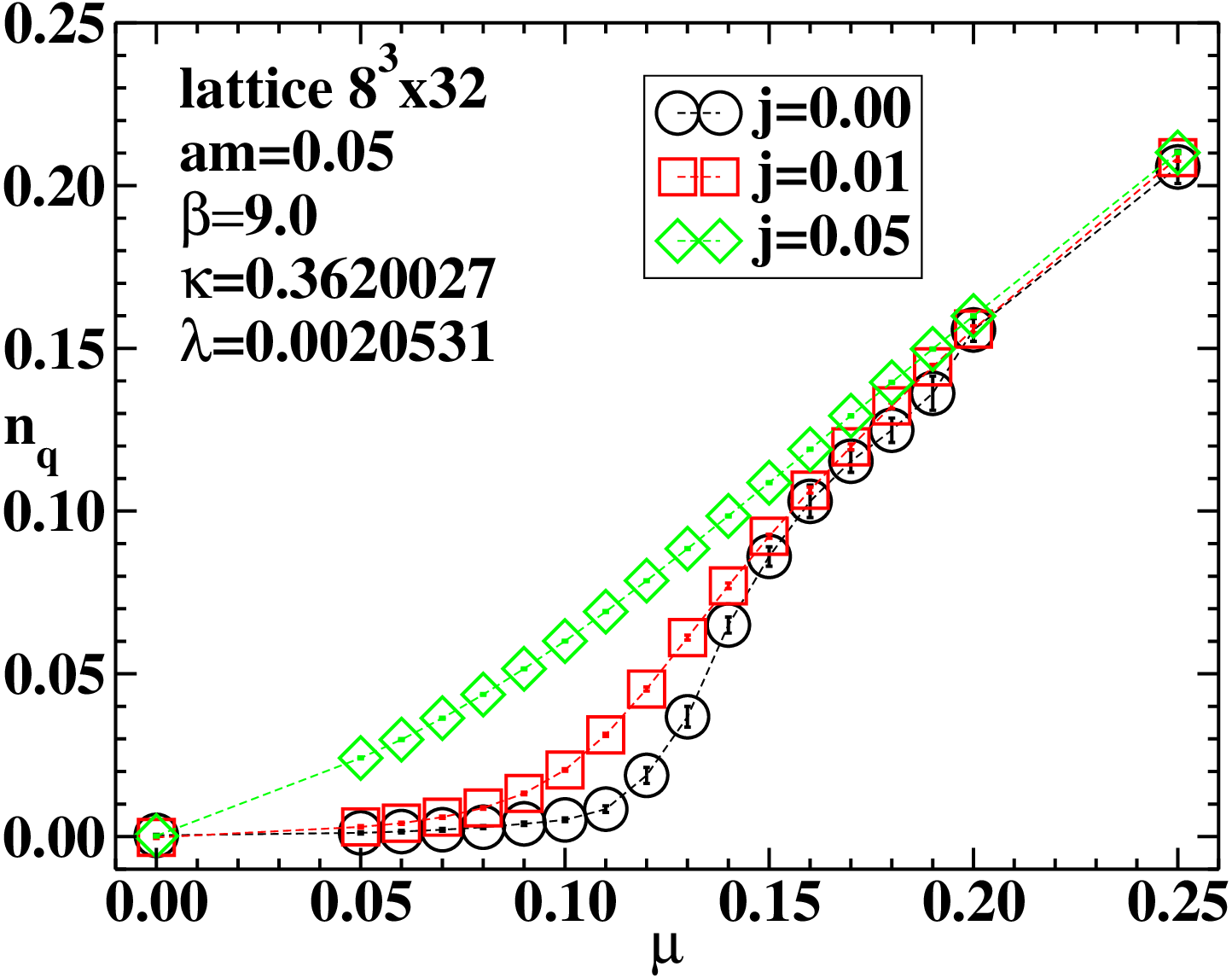, width=10.0cm}
\smallskip
\caption{$n_q$ vs. $\tilde\mu$ for various $j$ using the same
$\{\beta,\kappa,\lambda\}$.}
\label{fig:nqvsmu_kappa0.36}
\end{center}
\bigskip
\begin{center}
\epsfig{file=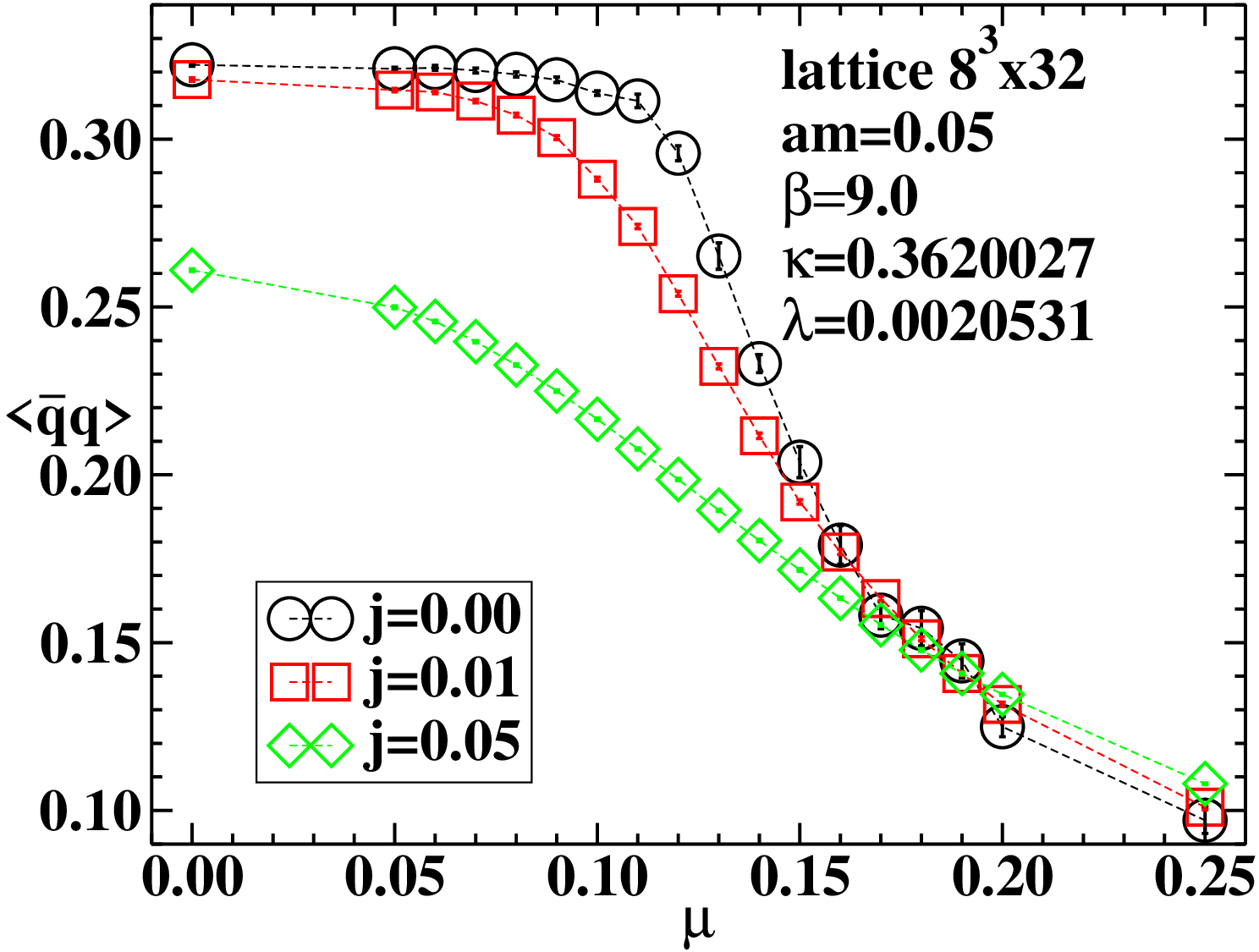, width=10.0cm}
\smallskip
\caption{$\langle\bar qq\rangle$ vs. $\tilde\mu$ for various $j$ using the same
$\{\beta,\kappa,\lambda\}$.}
\label{fig:pbpvsmu_kappa0.36}
\end{center}
\end{figure}
\begin{figure}
\begin{center}
\epsfig{file=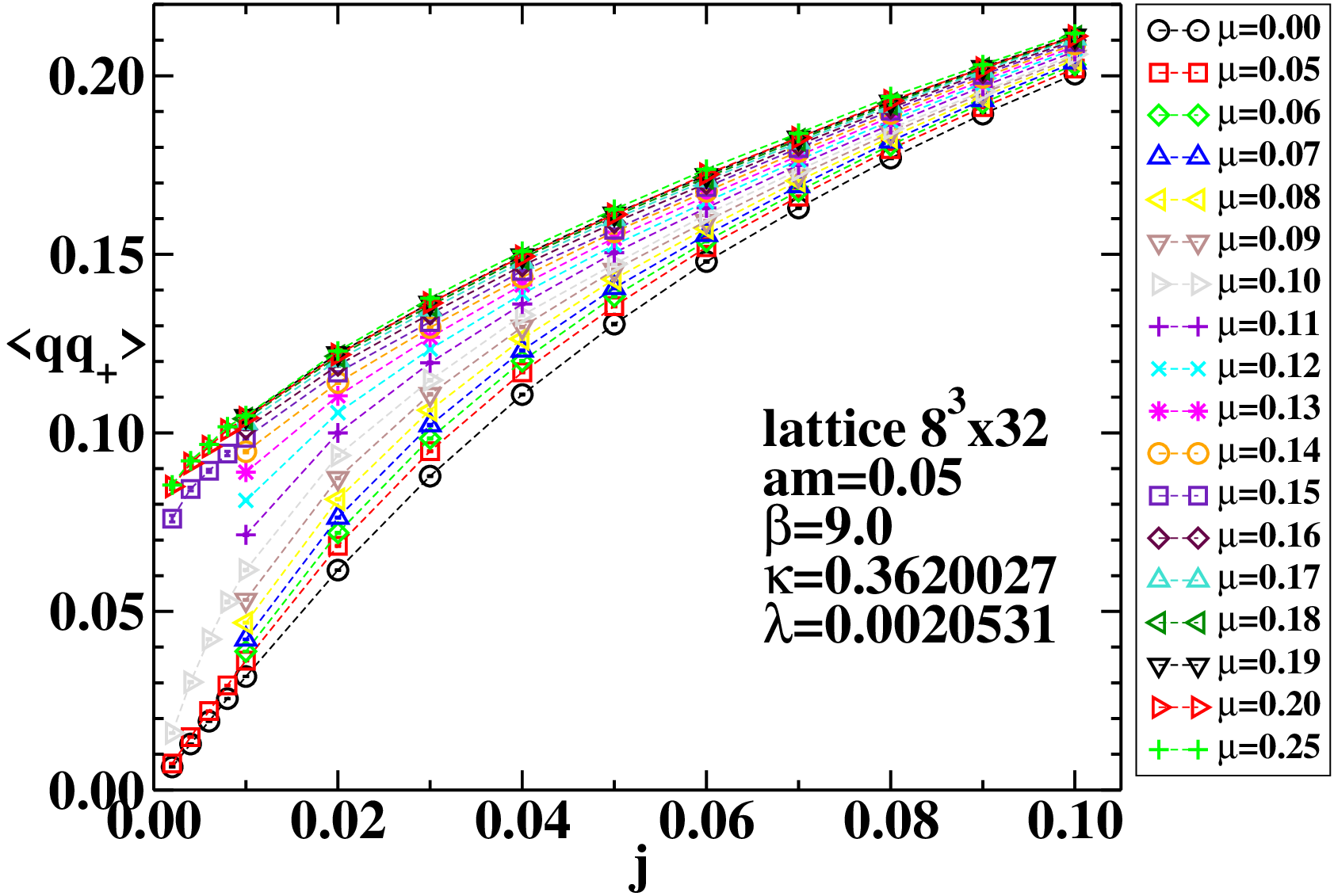, width=12.0cm}
\smallskip
\caption{$\langle qq_+\rangle$ vs. $j$ for various $\tilde\mu$ using the same
$\{\beta,\kappa,\lambda\}$.}
\label{fig:qqvsj_kappa0.36}
\end{center}
\bigskip
\begin{center}
\epsfig{file=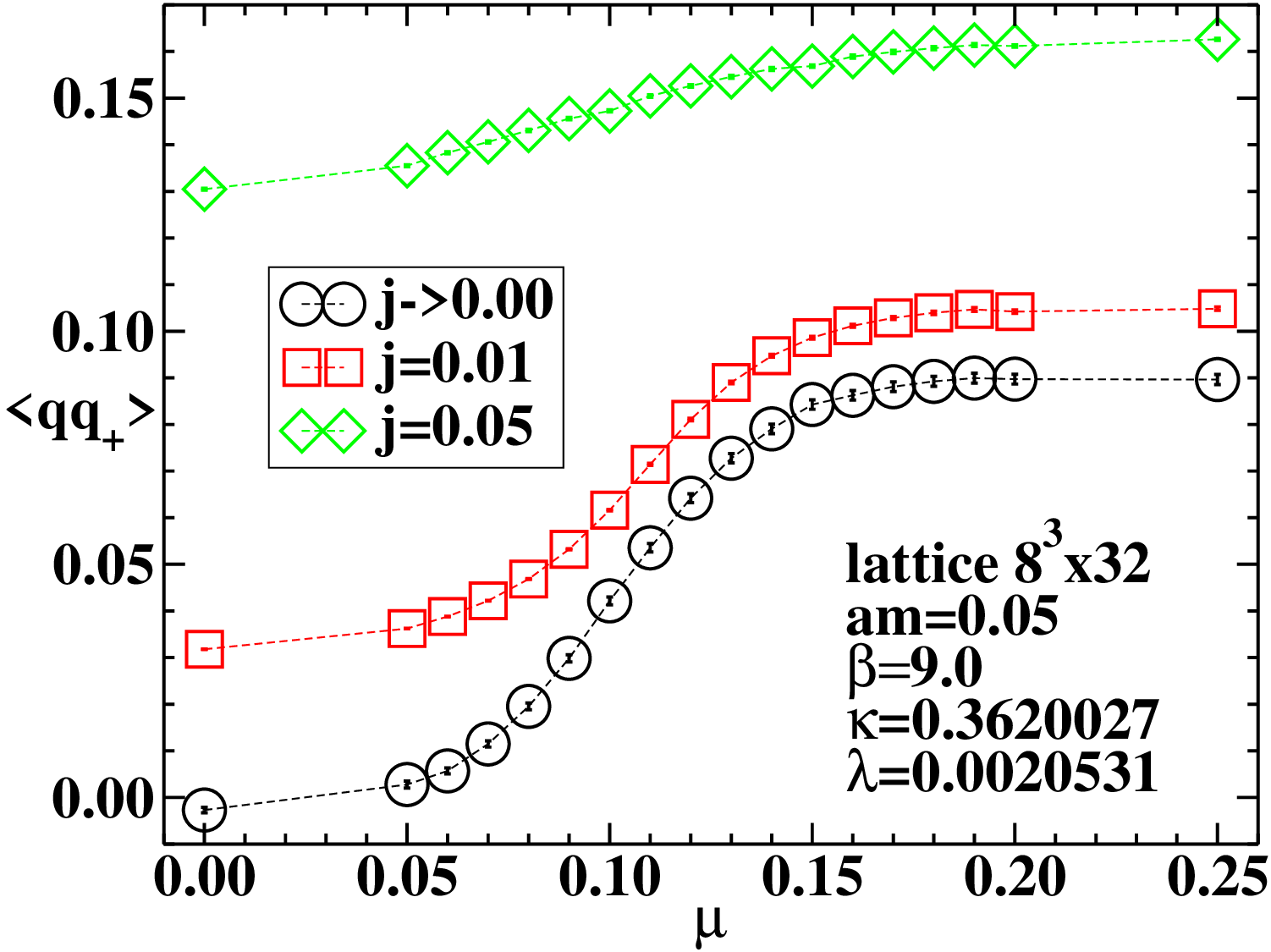, width=10.0cm}
\smallskip
\caption{$\langle qq_+\rangle$ vs. $\tilde\mu$ for various $j$ using the same
$\{\beta,\kappa,\lambda\}$.}
\label{fig:qqvsmu_kappa0.36}
\end{center}
\end{figure}

In Figs.~\ref{fig:nqvsmu_kappa0.36} and \ref{fig:pbpvsmu_kappa0.36} we plot
quark density $n_q$ and chiral condensate $\langle\bar qq\rangle$ as functions
of $\mu$ for various $j$, this time on a $8^3\times32$ lattice, with
$0\leq\mu/T\leq8$. 
The bare quark mass throughout this study was set to $ma=0.05$.
To confirm freedom from finite volume effects we also performed 
test simulations with $L_s=4,8,16,32$ and $L_t=16,32,64$.
There is a transition at $\tilde\mu\approx0.12$, becoming more abrupt as
$j\to0$, from a phase with
$n_q=0$, $\langle\bar qq\rangle$ constant to one in which $n_q$ increases
approximately linearly with $\mu$ and
$\langle\bar qq\rangle\propto\mu^{-2}$. This is in complete accordance with the
scenario described by $\chi$PT in which as $\mu$ increases at $T\approx0$
there is a transition at $\mu_c=m_\pi/2$ from the vacuum to a weakly-interacting
Bose gas formed from scalar diquarks (Cf. Figs. 4 and 5 of Ref.~\cite{KSTVZ}).
The diquarks are supposed to Bose-condense to form a superfluid condensate;
on a finite system this must be checked at $j\not=0$
using the $\langle qq_+\rangle$ observable of 
(\ref{eq:qq}). Fig.~\ref{fig:qqvsj_kappa0.36} shows a compilation of 
$\langle qq_+\rangle$ data as a function of $j$ for $\mu$ rising from
zero up to $\tilde\mu=0.25$, the condensate increasing monotonically with $\mu$.
To determine the nature of the ground state
an extrapolation $j\to0$ is needed. We have used
a cubic polynomial for data with $0.02\leq ja\leq0.1$, which may result in some
systematic uncertainty in the immediate neigbourhood of the transition, but
Fig.~\ref{fig:qqvsmu_kappa0.36} confirms that once again there is an abrupt 
change of behaviour in the order parameter at $\mu\approx m_\pi/2$, and
that the high-$\mu$ phase is superfluid.

Next we explored a parameter set corresponding to a smaller scalar
``stiffness'' 
by changing to $\kappa=0.1$ while keeping $ma=0.05$ 
-- naively following Table~\ref{tab_params}
suggests this corresponds to a huge value of $T/T_c$, ie. taking us further into
the deconfined phase of the hot gauge theory.  Of course, whether 
DR-based concepts
remain valid for the reconstructed theory must be addressed empirically.
\begin{figure}
\begin{center}
\epsfig{file=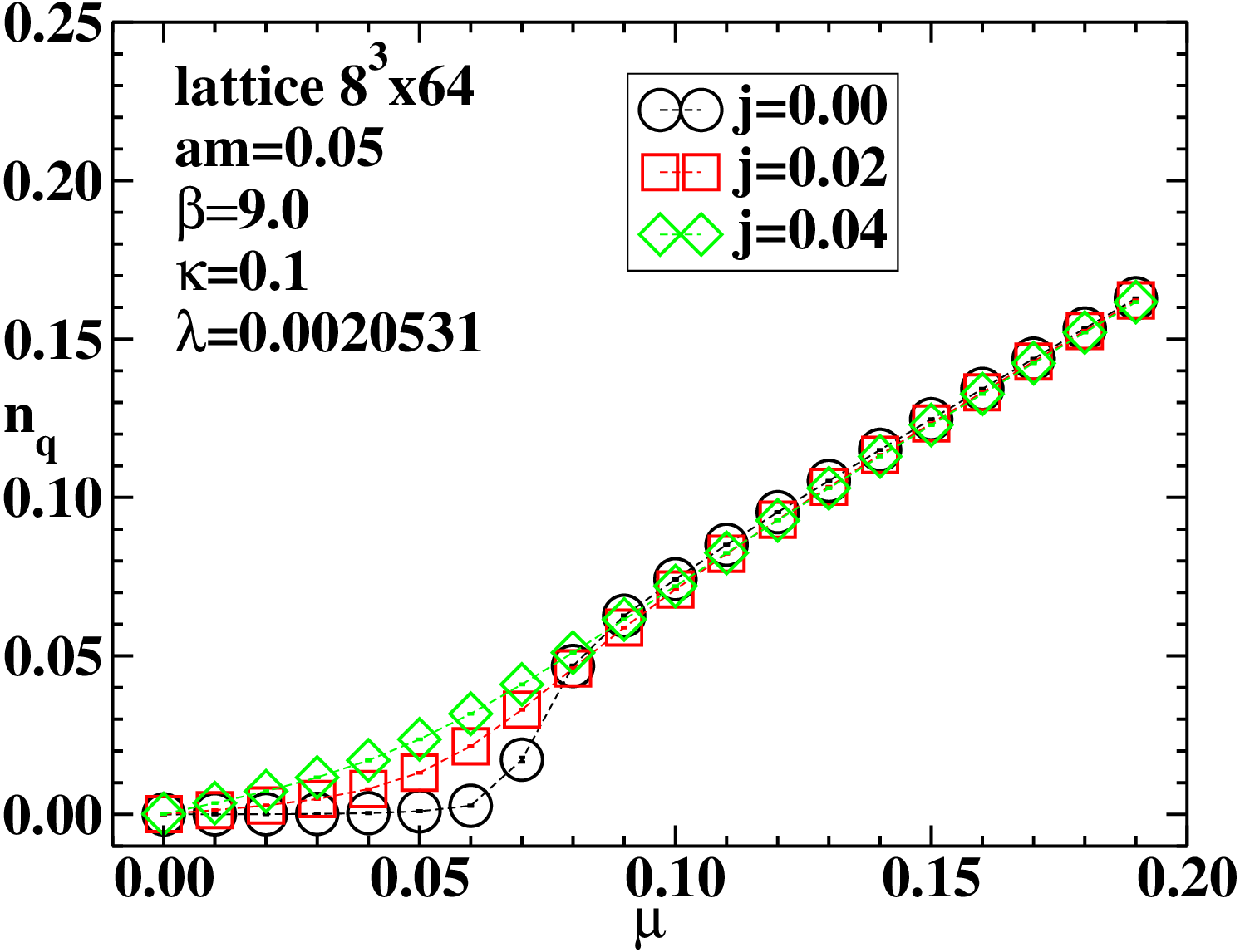, width=10.0cm}
\smallskip
\caption{$n_q$ vs. $\tilde\mu$ for various $j$ at $\beta=9.0$, $\kappa=0.1$,
$\lambda=0.0020531$.}
\label{fig:nqvsmu_kappa0.10}
\end{center}
\bigskip
\begin{center}
\epsfig{file=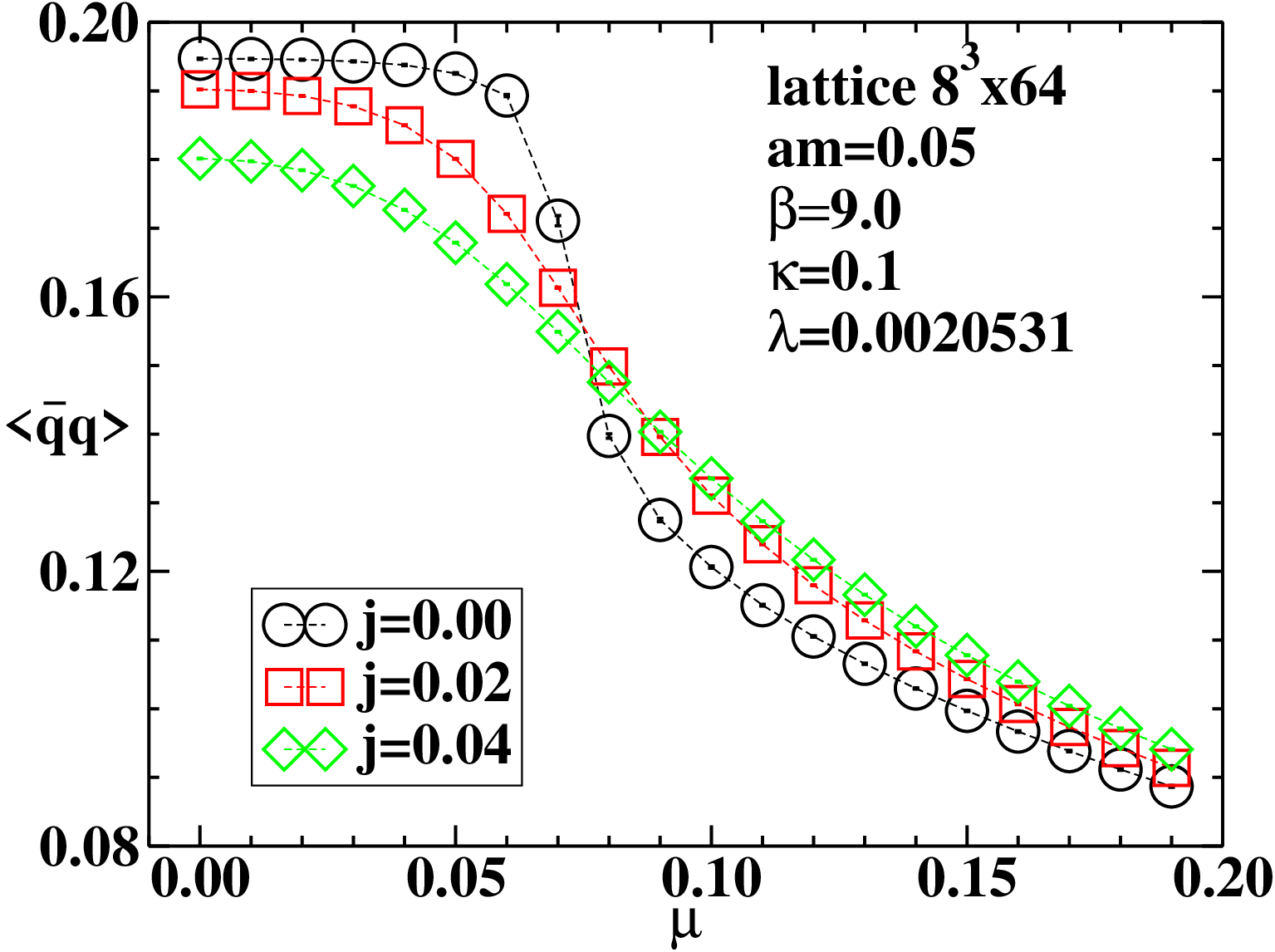, width=10.0cm}
\smallskip
\caption{$\langle\bar qq\rangle$ vs. $\tilde\mu$ for various $j$ using the same
$\{\beta,\kappa,\lambda\}$.}
\label{fig:pbpvsmu_kappa0.10}
\end{center}
\end{figure}
This time we used a volume $8^3\times64$ for the bulk observables
and at $\mu=0$ determined the pion mass $m_\pi a_t=0.1377(1)$ on $8^3\times
128$, and $m_\pi a_t=0.1423(4)$ on $16^3\times64$, 
showing that the finite volume
error is now roughly 3\%.
Figs.~\ref{fig:nqvsmu_kappa0.10} and \ref{fig:pbpvsmu_kappa0.10} show
respectively $n_q$ and $\langle\bar qq\rangle$ as functions of $\mu$ 
for $0\leq\mu/T\lapprox12$ in the
same format as previously. It is noteworthy that for $\mu>m_\pi/2$
$n_q(\mu)$ is numerically very
similar to the values found at $\kappa=0.3620027$, whereas for $\mu<m_\pi/2$
the chiral condensate $\langle\bar qq\rangle$ is significantly smaller,
indicative of a weaker quark -- anti-quark binding at this smaller $\kappa$.
As before, there is a clear discontinuity in the
observables' behaviour at $\mu_c\simeq m_\pi/2$, and the general picture is
qualitatively very similar, suggesting that the $\chi$PT
scenario is still applicable. Diquark binding is now also 
much weaker, however, as 
shown by the $\langle qq_+\rangle$ data of
Fig.~\ref{fig:qqvsmu_kappa0.10}.
\begin{figure}
\begin{center}
\epsfig{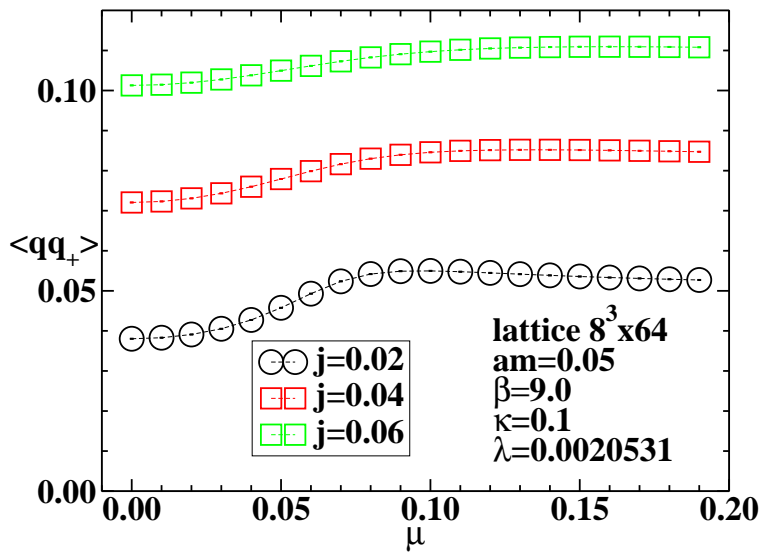}
\smallskip
\caption{$\langle qq_+\rangle$ vs. $\tilde\mu$ for various $j$ using the same
$\{\beta,\kappa,\lambda\}$.}
\label{fig:qqvsmu_kappa0.10}
\end{center}
\end{figure}
Note that the source values $j$ are greater than those of
Fig.~\ref{fig:qqvsmu_kappa0.36}, but the corresponding $\langle qq_+\rangle$ are
smaller in magnitude.
Since the curvature of the data as a function of $j$ is quite pronounced, a
reliable extrapolation $j\to0$ is impracticable on this system size. Indeed,
inspection by eye would suggest a linear extrapolation would yield $\langle
qq_+\rangle\simeq0$ for all $\mu$; however, the $ja=0.02$ data do manifest some
significant variation at $\mu\approx m_\pi/2$, suggestive that weak
symmetry-breaking persists for $\mu>\mu_c$.

\begin{figure}
\begin{center}
\epsfig{file=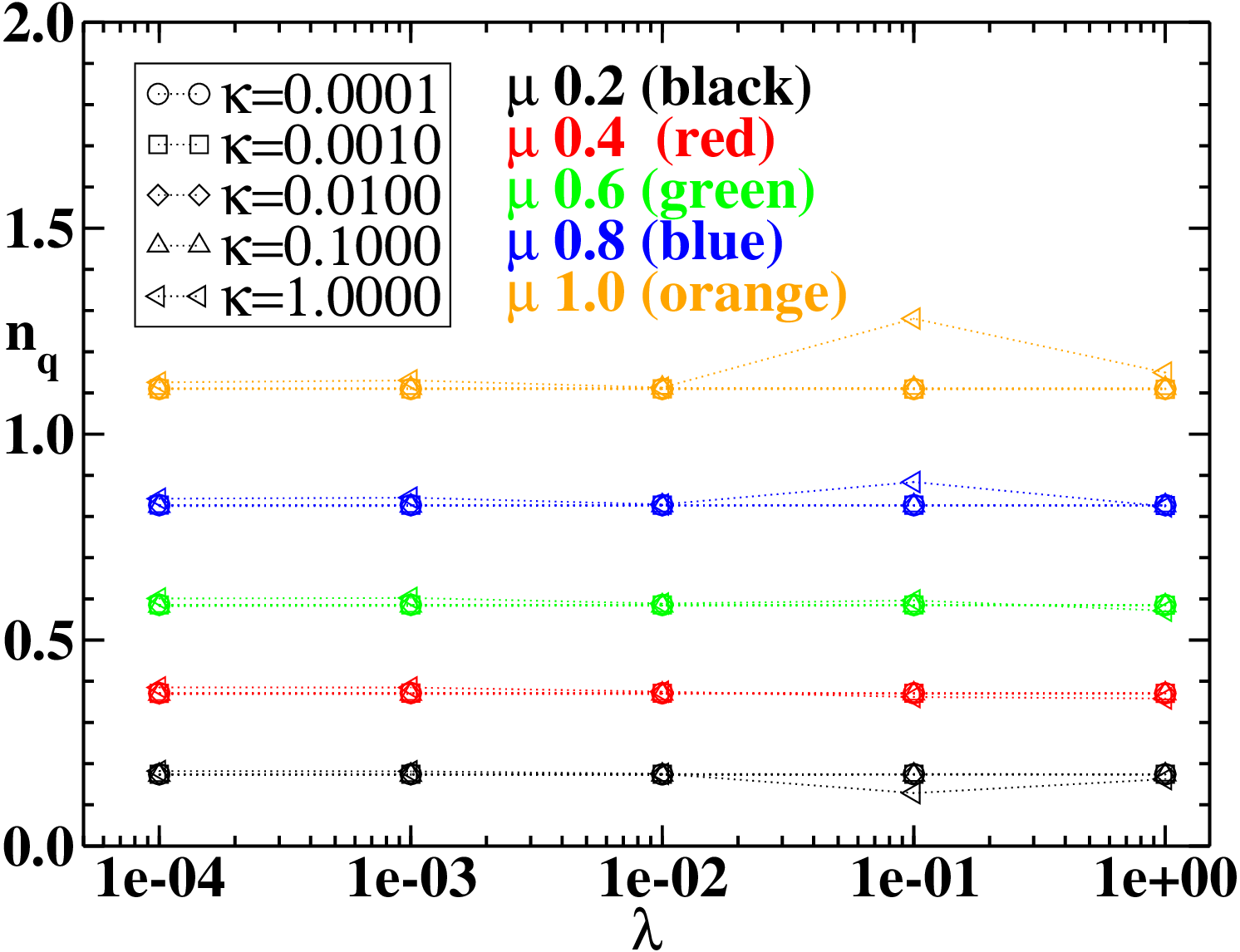, width=10.0cm}
\smallskip
\caption{$n_q$ vs. $\lambda$ for various $\tilde\mu$, $\kappa$.
Note $n_q$ increases systematically with $\mu$.}
\label{fig:nqvslambda_mukap}
\end{center}
\bigskip
\begin{center}
\epsfig{file=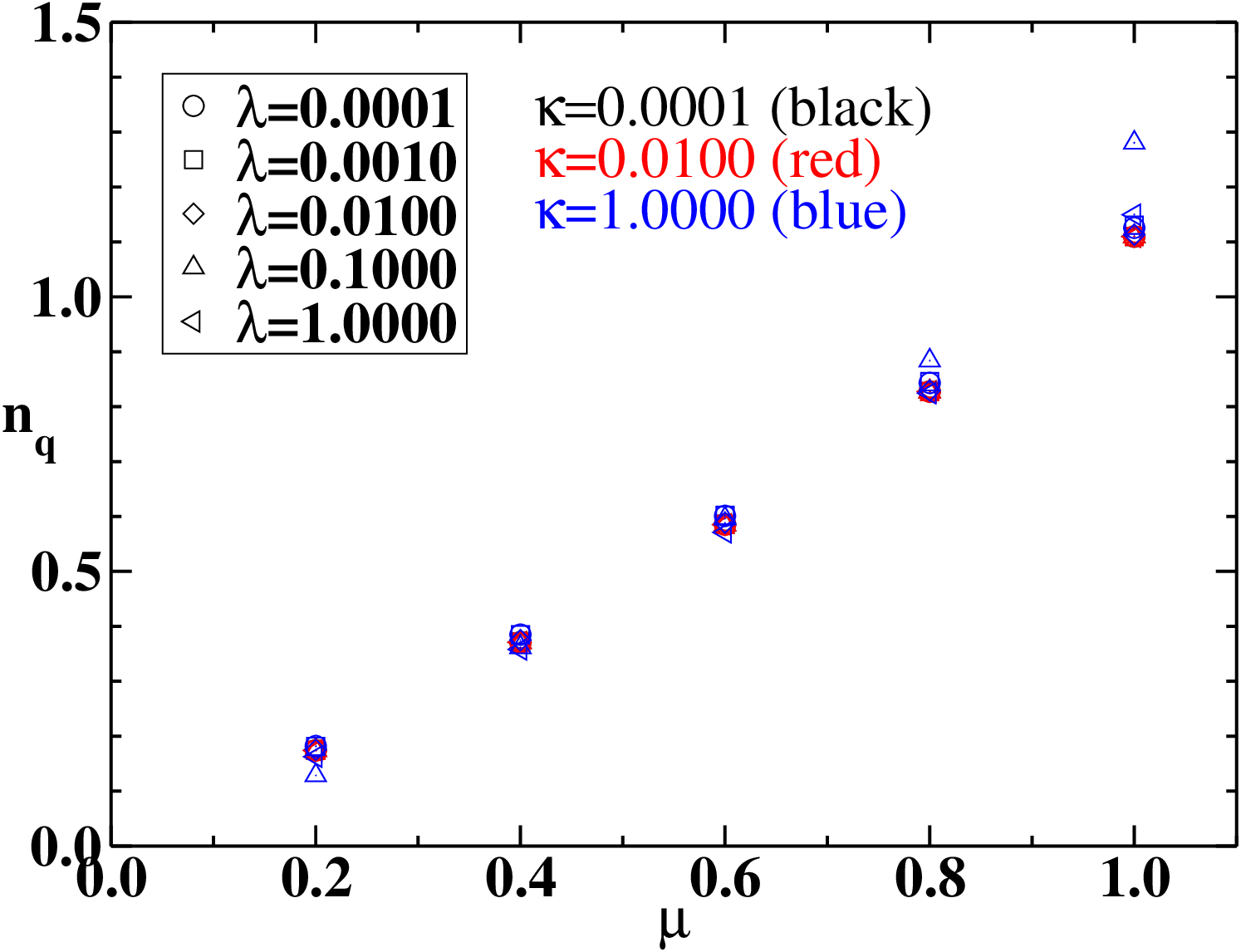, width=10.0cm}
\smallskip
\caption{$n_q$ vs. $\tilde\mu$ for various $\kappa$, $\lambda$.}
\label{fig:nqvsmu_lamkap}
\end{center}
\end{figure}

It is disappointing that we have found no qualitative change in physics as the
parameters are varied -- recall that the $\chi$PT model which describes the
results reasonably well is based on the assumption of confinement, or at least
on the presence of very tightly bound diquark states in the spectrum.
To explore the parameter space more widely we focussed on a single observable,
$n_q$, and scanned the $(\kappa,\lambda)$ plane on $8^3\times16$
at five different values of
$\mu$ with $\beta=9.0$, $ma=0.05$ and $ja=0.01$.
The results are summarised in Figs.~\ref{fig:nqvslambda_mukap} and
\ref{fig:nqvsmu_lamkap}. Fig.~\ref{fig:nqvslambda_mukap} shows that except
for $\lambda=0.1$ the results for fixed $\mu$ are practically independent 
of $\kappa$ (shown by the overlapping symbols) and of $\lambda$ (shown by the 
horizontal lines). Fig.~\ref{fig:nqvsmu_lamkap} shows that $n_q$ increases
linearly with $\mu$ over a wide region of parameter space, as it does for 
$\mu>\mu_c$ in Figs.~\ref{fig:nqvsmu_kappa0.36},\ref{fig:nqvsmu_kappa0.10}.
This approximate linear behaviour is once again a prediction of $\chi$PT
\cite{TCQCD,TCQCD2,KSTVZ}, 
and is to be contrasted with the $n_q\propto\mu^3$ behaviour
expected of a deconfined theory where baryons can be identified with 
degenerate quark states occupying a Fermi sphere of radius $k_F\approx\mu$.
The absence of this scaling is a further reason to conclude that the
reconstructed model does not describe deconfined
physics.

\subsection{Bosonic Spectrum}

Next we report on bound state spectroscopy. Conventionally, $q\bar
q$ bound states are referred to as mesons, and $qq$, $\bar q\bar q$ as diquark
baryons
and anti-baryons respectively. In a medium with spontaneously broken baryon
number symmetry, however, excitations need not have a well-defined baryon
number. An appropriate set of states to look at,
together with gauge-invariant 
interpolating operators expressed in terms of staggered fermion fields,
is as follows:
\begin{eqnarray}
\mbox{pion} && \bar\chi_x\varepsilon_x\chi_x; \label{eq:pion}\\
\mbox{scalar} && \bar\chi_x\chi_x; \\
\mbox{higgs} && {1\over2}
(\chi^t_x\tau_2\chi_x+\bar\chi_x\tau_2\bar\chi^t_x)\equiv qq_{+x}; 
\label{eq:higgs} \\
\mbox{goldstone} && {1\over2}
(\chi^t_x\tau_2\chi_x-\bar\chi_x\tau_2\bar\chi^t_x)\equiv qq_{-x}.
\label{eq:goldstone}
\end{eqnarray}
Here the phase $\varepsilon_x=(-1)^{x_0+x_1+x_2+x_3}$, and the Pauli matrix 
$\tau_2$ acts on color indices. Pion and scalar states are related via the
U(1)$_\varepsilon$ global symmetry $\chi\mapsto e^{i\alpha\varepsilon}\chi$,
$\bar\chi\mapsto\bar\chi e^{i\alpha\varepsilon}$. Analogous to
chiral symmetry for continuum spinors, this is an exact symmetry of the 
action (\ref{eq:fermaction}) in the limit $m\to0$. In a phase with spontaneously
broken chiral symmetry, the scalar is massive, and the pion a Goldstone mode,
becoming massless as $m\to0$. Similarly, ``higgs'' and ``goldstone'' diquark 
states are related via the U(1)$_B$ baryon number rotation
$\chi\mapsto e^{i\beta}\chi$, $\bar\chi\mapsto\bar\chi e^{-i\beta}$, an exact
symmetry of (\ref{eq:fermaction},\ref{eq:dsource}) in the limit $j\to0$.
In a superfluid phase with $\langle qq_+\rangle\not=0$, the higgs is massive,
and the goldstone massless in the limit $j\to0$.

The boson correlators are constructed from the 
{\em Gor'kov}
propagator \cite{HW}
\begin{equation}
{\cal
G}_{xy}=\left(\matrix{\bar\jmath\tau_2&M(\mu)\cr
-M^{tr}(\mu)&j\tau_2\cr}\right)^{-1}_{xy}
=\left(\matrix{A_{xy}&N_{xy}\cr\bar N_{xy}&\bar A_{xy}}\right),
\label{eq:Gor'kov}
\end{equation}
where the $2\times2$ (in color space) matrices 
$N\sim\langle\chi_x\bar\chi_y\rangle$ and $A
\sim\langle\chi_x\chi_y\rangle$ are known as the {\em normal} and {\em
anomalous} parts respectively. On a finite volume $A\equiv0$ for $j=0$; 
$\lim_{j\to0}\lim_{V\to\infty}A\not=0$ signals particle-hole mixing resulting
from the breakdown of U(1)$_B$ symmetry, and hence superfluidity.
Due to SU(2) symmetries the only 
independent components of ${\cal G}$ are $\mbox{Re}N_{11}\equiv N$ and 
$\mbox{Im}A_{12}\equiv A$ and their barred counterparts, 
so that in practice all information from a single source 
can be extracted at $\mu\not=0$ using
just two matrix inversions per configuration. Each of the correlators $C(t)$
constructed from the forms (\ref{eq:pion}-\ref{eq:goldstone})
receives contributions from diagrams containing both $N$ and $A$-type
propagators; by construction, however, they remain symmetric under $t\mapsto-t$
even once $\mu\not=0$. Note also that in this quenched treatment contributions
to the mesons from disconnected $N$ loops and to the diquarks from disconnected
$A$ loops are neglected.

\begin{figure}
\begin{center}
\epsfig{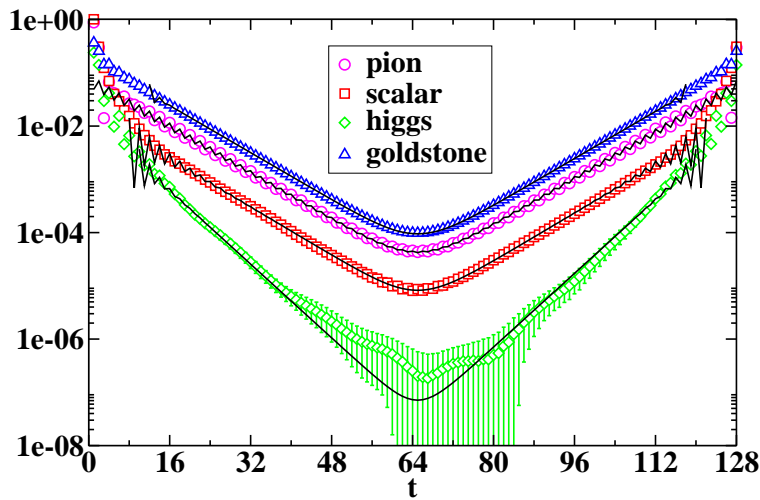}
\smallskip
\caption{Timeslice propagators of various bosonic bound states at 
$\tilde\mu=0.25$
on a $8^3\times128$ lattice at $\beta=9.0$, $\kappa=0.3620027$,
$\lambda=0.0020531$, $ma=0.05$, $ja=0.02$.}
\label{fig:props}
\end{center}
\end{figure}
Fig.~\ref{fig:props} shows all four timeslice correlators evaluated on a
$8^3\times128$ lattice with parameter set
$\beta=9.0$, $\kappa=0.3620027$ and 
$\lambda=0.0020531$.  The chemical potential $\tilde\mu=0.25$, 
ie. above the critical
$\tilde\mu_c$ required to enter the superfluid phase. 
The quark bare mass $ma=0.05$ and diquark source
$ja=\bar\jmath a=0.02$. All four channels yield clear signals for
single particle bound states,
the higgs being the noisiest. 

Like any meson constructed from staggered fermions,
the correlators in principle describe two states and must
be fitted using the form
\begin{equation}
C(t)=A[e^{-mt}+e^{-m(L_t-t)}]+B[e^{-Mt}+(-1)^te^{-M(L_t-t)}].
\label{eq:propfit}
\end{equation}
where $m$ and $M$ denote the masses of states with opposite parities.
Fits to (\ref{eq:propfit}), where positive, are shown by solid lines in 
Fig.~\ref{fig:props}.
In most cases we find $M\gg m$; however for $\mu>\mu_c$ the pion correlator 
has a distinct ``saw-tooth'' shape, and in fact the fit yields
$m_\pi>M_{b1}$, where $\pi$ denotes the usual pseudoscalar pion, and $b1$ a
state of opposite parity, which must therefore be scalar (note that the
logarithmic scale requires 
Fig.~\ref{fig:props} to plot $\vert C_\pi(t)\vert$).
\begin{figure}
\begin{center}
\epsfig{file=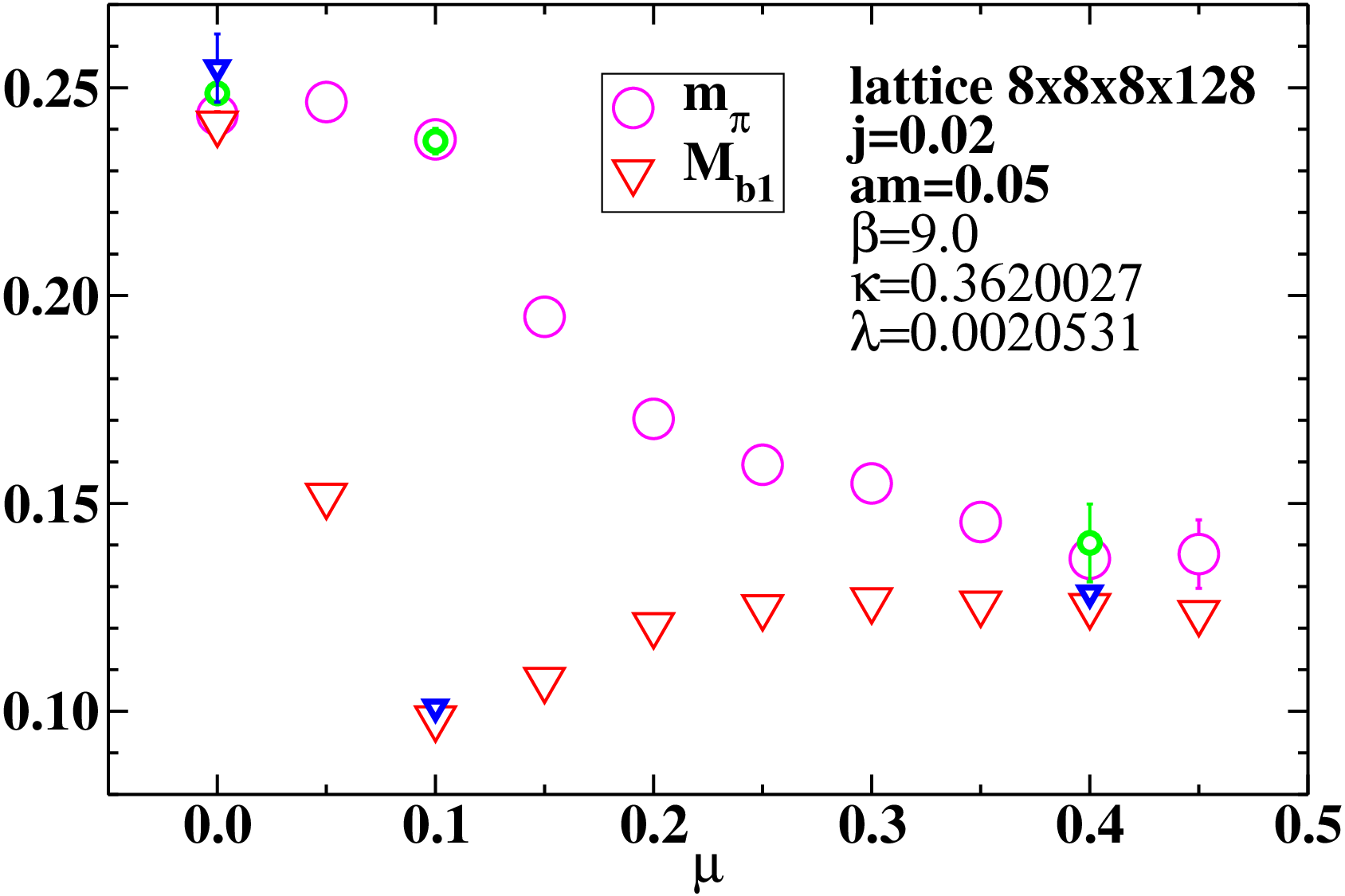, width=10.0cm}
\smallskip
\caption{Mass spectrum of the pion and its parity partner 
as a function of $\tilde\mu$. 
Smaller symbols denote data taken on $16^3\times64$.}
\label{fig:mpib1vsmu}
\end{center}
\bigskip
\begin{center}
\epsfig{file=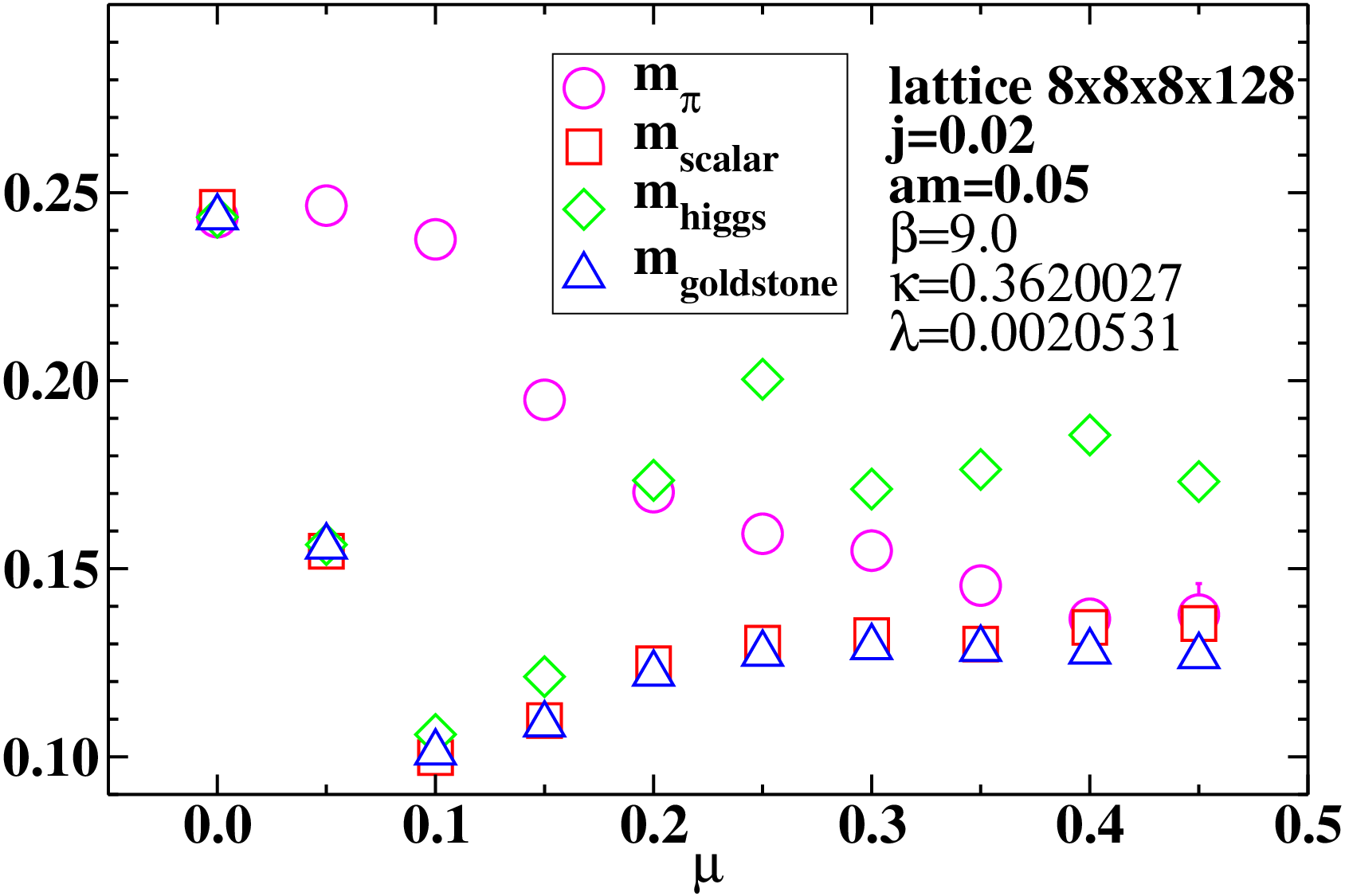, width=10.0cm}
\smallskip
\caption{Mass spectrum of various bosonic excitations as a function of 
$\tilde\mu$.}
\label{fig:bosons}
\end{center}
\end{figure}

In Fig.~\ref{fig:mpib1vsmu} we plot $m_\pi$ and $M_{b1}$ against $\mu$, and
in Fig.~\ref{fig:bosons} the corresponding spectrum for all the states
(\ref{eq:pion}-\ref{eq:goldstone}).  Fig.~\ref{fig:mpib1vsmu} also shows
$m_\pi$ and $M_{b1}$ as measured on a $16^3\times64$ system at some
representative values of $\mu$. 
Even for the lightest measured mass,
$M_{b1}^{-1}(\tilde\mu=0.9)\approx10<L_s$; 
while volume effects are
statistically significant, they have no impact on the qualitative trends we now
discuss. 
First note that all states are
approximately degenerate at $\mu=0$. The equality of pion, higgs and goldstone
correlators 
is guaranteed by SU(2) symmetry at $\mu=0$ \cite{HKLM}, but the degeneracy of
the scalar in the chirally-broken vacuum suggested by
Fig.~\ref{fig:pbpvsmu_kappa0.36} can only arise as a result of meson-diquark
mixing due to $j\not=0$.
Next,
note that the pion mass remains
constant for $\mu<\mu_c$, where as reviewed above it is a pseudo-Goldstone boson
associated with chiral symmetry breaking, 
and then falls once the superfluid phase is entered.
This is in accordance with the predictions of $\chi$PT for the so-called
``$P_S$'' state of a theory with Dyson index $\beta_D=4$ \cite{KSTVZ}, and has
also been observed in simulations with dynamical staggered quarks in the
fundamental representation of SU(2) \cite{TCQCD2}. Most of the other states, 
including the $b1$ in Fig.~\ref{fig:mpib1vsmu}, show a
much steeper decrease with $\mu$ for $\mu<\mu_c$, followed by a gentle rise to a
plateau at $ma\approx0.13$ in the superfluid phase $\mu>\mu_c$, 
precisely that
expected of the goldstone state expected in the superfluid phase with diquark
source $j\not=0$ (Cf. the ``$Q_I$'' state shown in Fig. 3 of \cite{KSTVZ}).
The exception is the 
higgs, which rises more steeply to become the heaviest state at large
$\mu$.  We conclude {\em(a)\/} the breaking of degeneracy between higgs
(\ref{eq:higgs}) and goldstone (\ref{eq:goldstone}) states is clear
supplementary evidence for the breaking of U(1)$_B$ symmetry in the superfluid
phase; 
{\em(b)\/} all states with $J^P=0^+$ including the $b1$ but {\em
except\/} the higgs have some projection onto the Goldstone state, regardless of
whether the original interpolating operator is mesonic or baryonic. 
It would be interesting to study this phenomenon as $j$ is varied.

\subsection{Fermionic Spectrum}

We have also studied the fermion spectrum, 
often in the context of condensed matter
called the
{\em quasiparticle} spectrum.
Since the Gor'kov propagator
${\cal G}$ is not gauge invariant we have to specify a gauge fixing
procedure. 
A feature of the quenched approach is that it permits large statistics to be
accumulated with relatively little CPU effort. This has enabled us for the
first time in a gauge theory context to study ${\cal G}$ at
$\mu\not=0$, by helping to overcome the sampling problems associated with 
gauge fixing.
We have experimented with two gauge choices:
{\em Unitary gauge\/} $\varphi\mapsto\varphi^\prime=
(0,0,\varphi_3^\prime)$, which is implemented {\em before} 
the reconstruction of the
4th dimension, and is unique up to a Z$_2$ factor, specified
by demanding $\varphi^\prime_3\geq0$; 
and {\em Coulomb gauge\/}, implemented by 
maximising $\sum_{xi}\mbox{tr}(U_{x,i}+U^\dagger_{x-\hat\imath,i})$, 
in an attempt to make the gauge fields as smooth as possible and
hence improve the signal-to-noise ratio.

\begin{figure}
\begin{center}
\epsfig{file=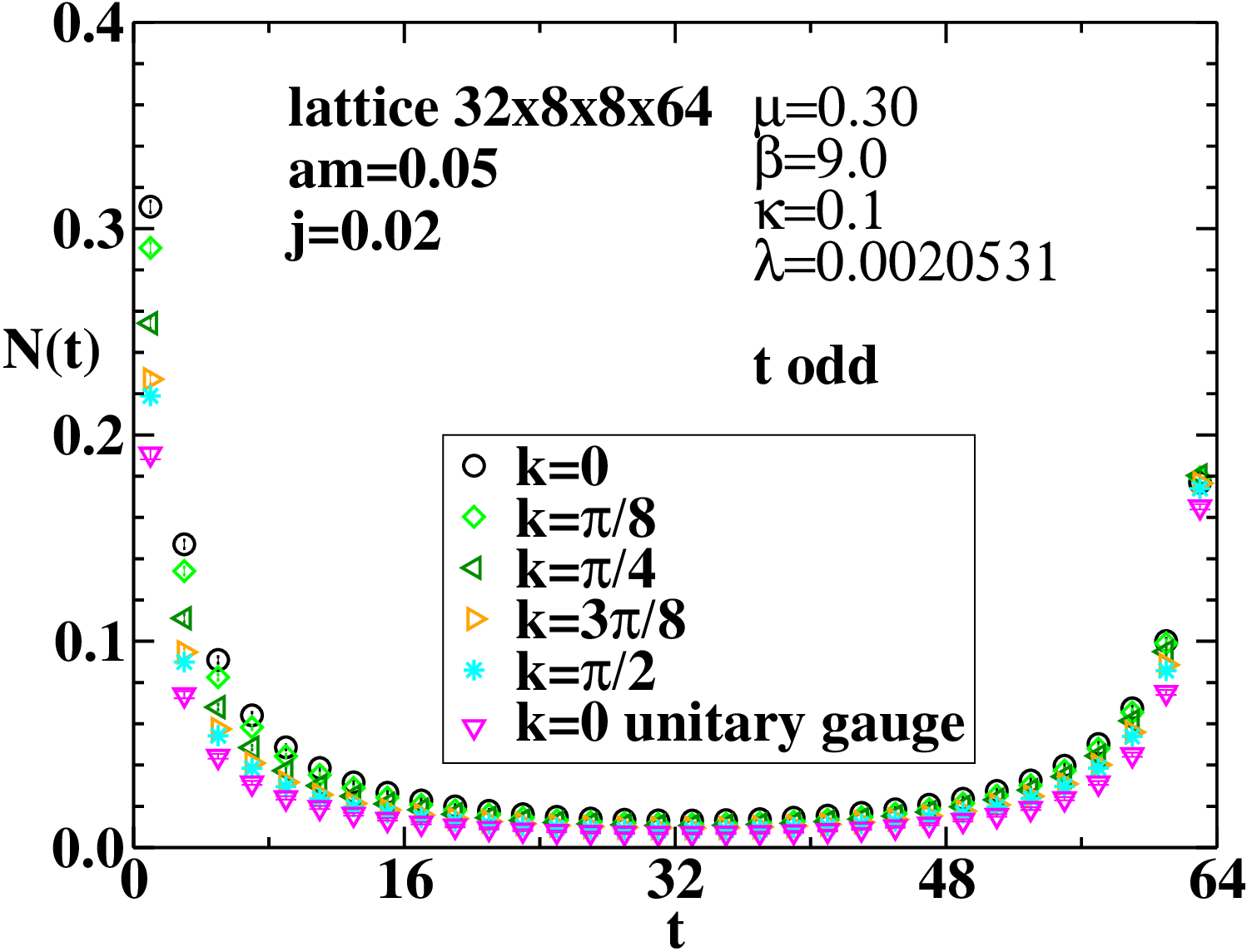, width=10.0cm}
\smallskip
\caption{Normal component of the quark propagator $N(t)$
for various momenta $k$ for $\mu>\mu_c$.}
\label{fig:propn_k0.1}
\end{center}
\bigskip
\begin{center}
\epsfig{file=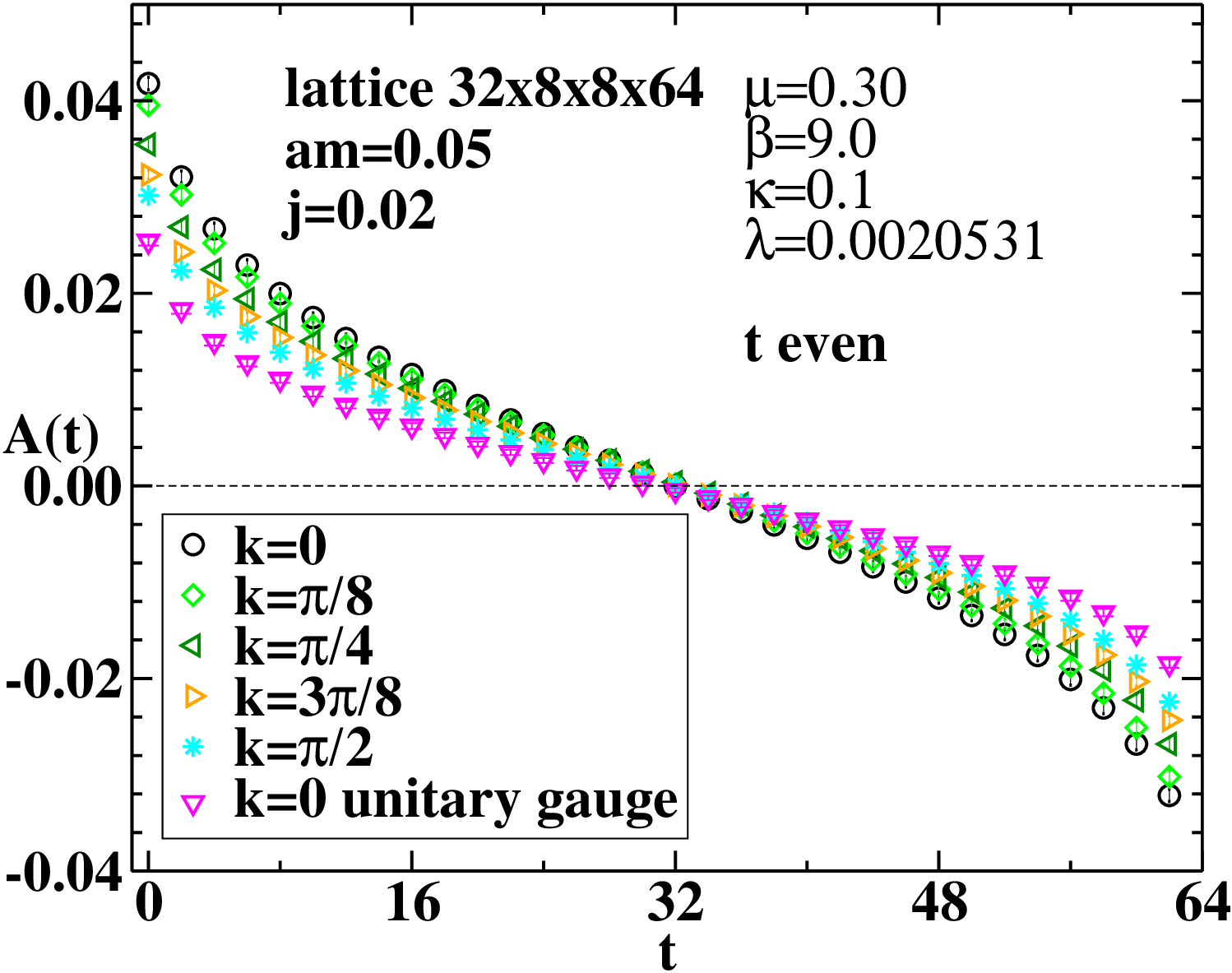, width=10.0cm}
\smallskip
\caption{Anomalous component of the quark propagator $A(t)$
for various momenta $k$ for $\mu>\mu_c$.}
\label{fig:propa_k0.1}
\end{center}
\end{figure}
In Figs.~\ref{fig:propn_k0.1} and ~\ref{fig:propa_k0.1} we plot respectively 
the normal and anomalous fermion timeslice propagators on a
$32\times8^2\times64$ lattice at $\beta=9.0$, $\kappa=0.1$, $\lambda=0.0020531$,
$ma=0.05$, $ja=0.0.2$
and $\tilde\mu=0.3$, the last value chosen to ensure $\mu>\mu_c$. 
These plots result from an analysis of 800 independent configurations.
As discussed in
Ref.~\cite{HW}, the properties of staggered lattice fermions are such that in a
phase with approximate chiral symmetry, as expected for $\mu>\mu_c$ from
Fig.~\ref{fig:pbpvsmu_kappa0.10}, the numerically important contributions to 
$N(t)$ are from $t$ odd, and to $A(t)$ from $t$ even, and only these
points are plotted. Most of the data 
was obtained using Coulomb gauge and differing values of the momentum
$k_xa\in\{0,{\pi\over16},\ldots,{\pi\over2}\}$, though the $k_x=0$ data taken in
unitary gauge are shown for comparison. Two features to note are that the 
Coulomb data is roughly twice as large as the unitary data reflecting an
enhanced signal as anticipated above,
and that there is little variation with $k_x$. 

In the NJL model
the quasiparticle propagator can be successfully fitted using the
forms~\cite{HW} 
\begin{eqnarray}
N(t)=Pe^{-E_Nt}+Qe^{-E_N(L_t-t)},\label{eq:fitN}\\
A(t)=R[e^{-E_At}-e^{-E_A(L_t-t)}].\label{eq:fitA}
\end{eqnarray}
where for $\mu\not=0$ there is no reason to expect $P=Q$, but for a well-defined
quasiparticle state the equality $E_N=E_A$ should hold.
Note that the anomalous amplitude 
$R\not=0$ is evidence for baryon number symmetry breaking, often in a condensed
matter context called ``particle-hole mixing''. 
\begin{table}[t]
\begin{center}
\begin{tabular}{|c|llll|}
\hline
          & Coulomb gauge  & $\chi^2$/dof &  Unitary gauge  & $\chi^2$/dof\\
\hline
$E_N$    & 0.1244(7) & 202 & 0.1651(17)  & 182 \\
$E_A$    & 0.0674(13) & 3.4 & 0.0696(15) & 5.9  \\
\hline
\end{tabular}
\caption{Fitted mass values to the data of Figs.~\ref{fig:propn_k0.1} and
\ref{fig:propa_k0.1} for $k_x=0$. \label{tab:massfits}}
\end{center}
\end{table}
Fits to the data of Figs.~\ref{fig:propn_k0.1} and \ref{fig:propa_k0.1} are
shown in Table~\ref{tab:massfits}, and it should be noted that only the
anomalous channel fits produced an acceptable $\chi^2$. 
\begin{figure}
\begin{center}
\epsfig{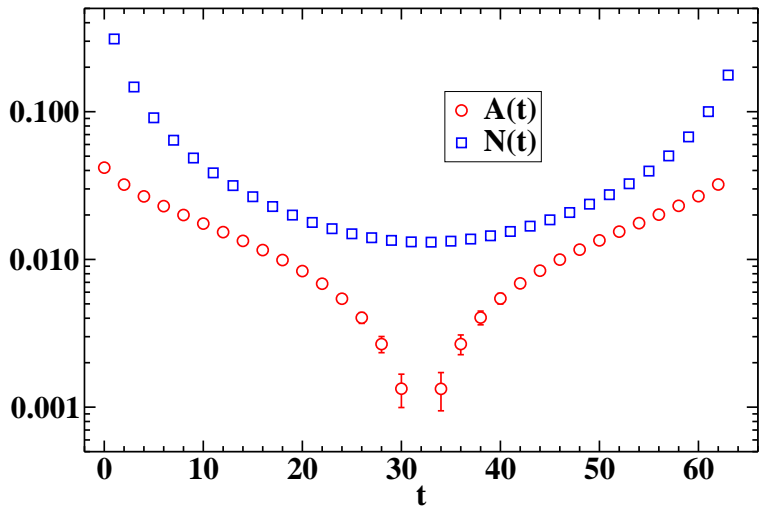}
\smallskip
\caption{Comparison of the $k_x=0$ data of Figs.~\ref{fig:propn_k0.1} and
\ref{fig:propa_k0.1} on a logarithmic scale.}
\label{fig:logscale}
\end{center}
\end{figure}
Some insight can be gained from Fig.~\ref{fig:logscale}, which compares the
Coulomb gauge normal and anomalous propagators at $k_x=0$ on a logarithmic
scale; while the fit (\ref{eq:fitA}) for $A(t)$ looks plausible, 
the normal channel never settles to a well-defined quasiparticle pole, even 
with $L_t=64$. 
Moreover only the anomalous channel shows any
evidence of gauge independence in Tab.~\ref{tab:massfits}. 
The value of $E_A$ obtained is very close to
$m_\pi/2$, indicating that at this value of $\kappa$ the pion is a weakly bound
state.
Another striking feature of the data is the
approximate forwards-backwards symmetry of $N(t)$, implying $P\simeq Q$. 

\begin{figure}
\begin{center}
\epsfig{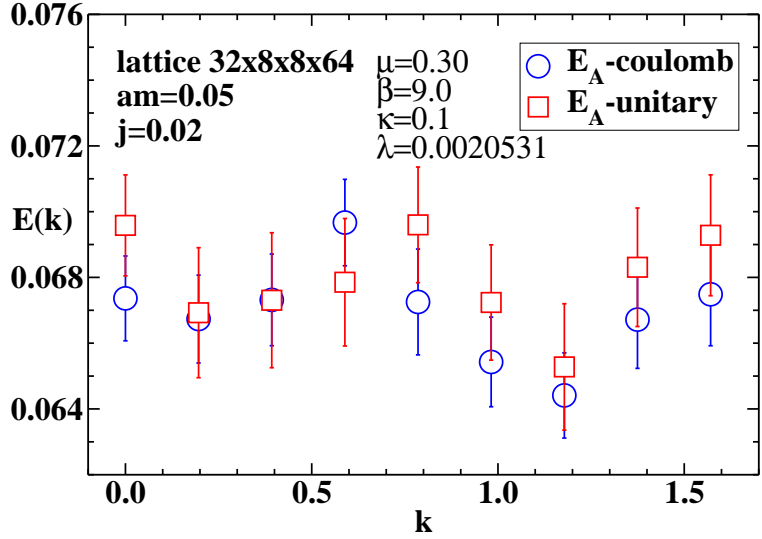}
\smallskip
\caption{Dispersion relations $E_N(k)$ and $E_A(k)$.}
\label{fig:dispersion}
\end{center}
\end{figure}
Fig.~\ref{fig:dispersion} shows the dispersion relations $E_A(k_x)$ for data
taken on a $32\times8^2\times64$ lattice at the parameter values shown 
(we were unable to obtain satisfactory fits to (\ref{eq:fitN}) to extract $E_N$
for $k_x\not=0$).
It confirms that the quasiparticle excitation
energies are $k$-independent. 
This should be contrasted with the findings of
\cite{HW}, where a lattice study of the NJL model using identical formalism
found $E(k)$ exhibiting a pronounced minimum at $k\approx k_F\approx\mu$, 
the Fermi momentum,
evidence that the NJL
model has a well-defined Fermi surface with $k_F\approx\mu$. One motivation for
our study was to investigate to what extent the concept of a Fermi surface,
which is not strictly
gauge invariant, can be put on a firm empirical footing in a gauge
theory. Fig.~\ref{fig:dispersion} shows no evidence for a Fermi surface.

\begin{figure}
\begin{center}
\epsfig{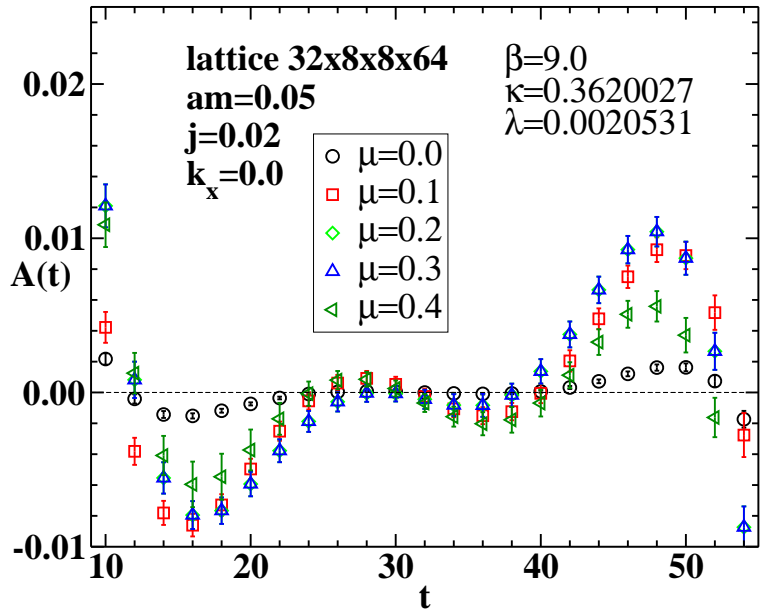}
\smallskip
\caption{Close-up of $A(t)$ for $\kappa=0.3620027$ for various $\tilde\mu$.}
\label{fig:wiggles}
\end{center}
\end{figure}
\begin{figure}
\begin{center}
\epsfig{file=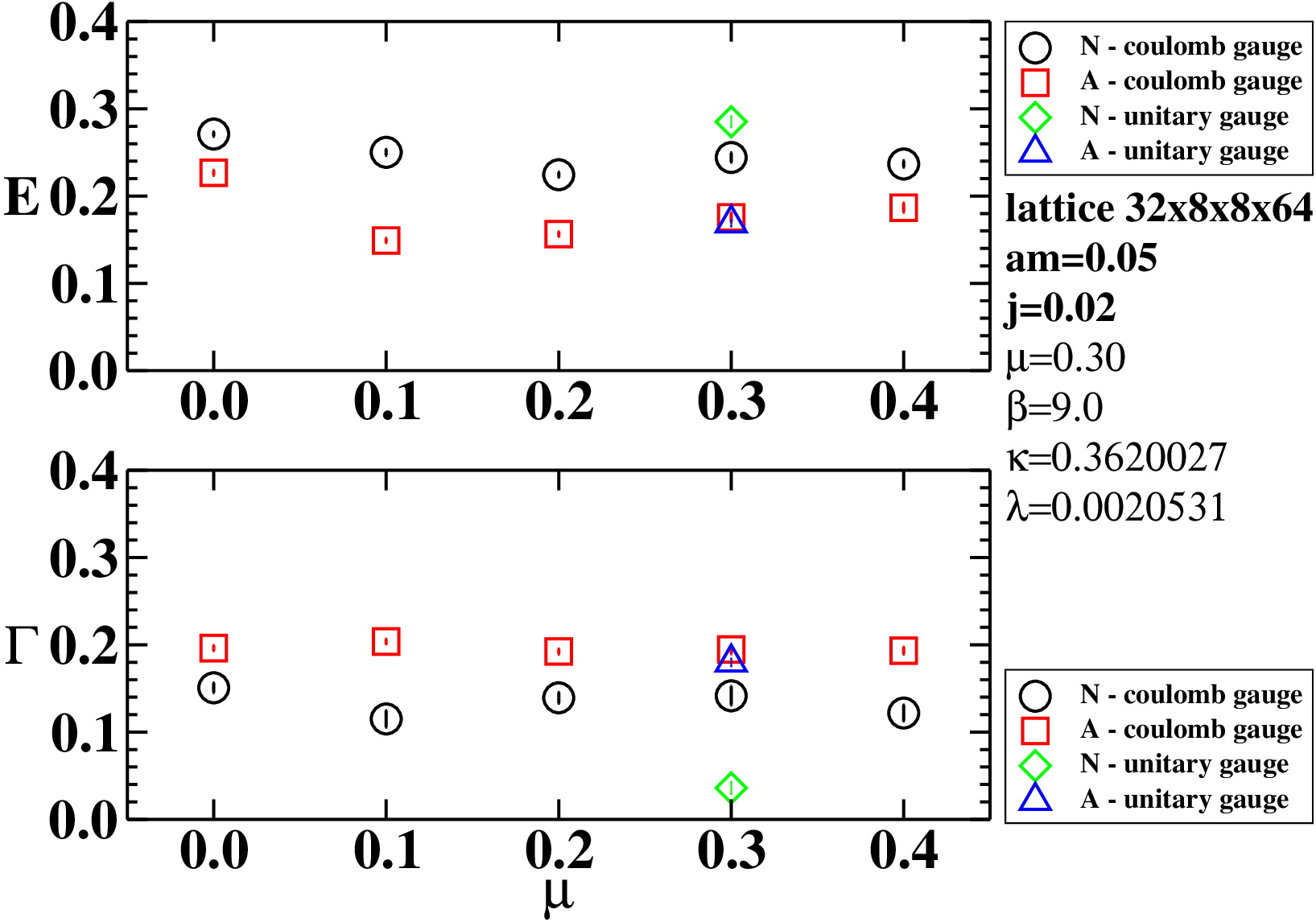, width=10.0cm}
\smallskip
\caption{$E$ and $\Gamma$ vs. $\tilde\mu$ for $\kappa=0.3620027$
in both normal and anomalous channels.}
\label{fig:mgamma}
\end{center}
\bigskip
\begin{center}
\epsfig{file=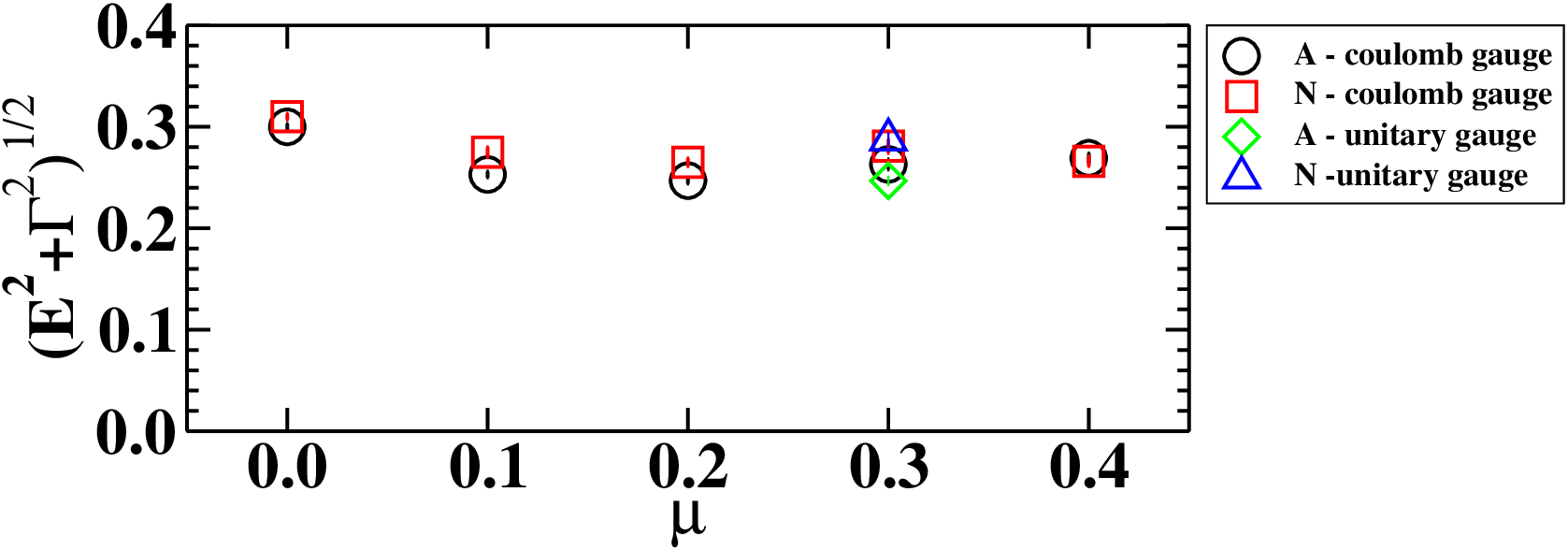, width=10.0cm}
\smallskip
\caption{$\sqrt{E^2+\Gamma^2}$ vs. $\tilde\mu$ for $\kappa=0.3620027$
in both normal and anomalous channels.}
\label{fig:modmgamma}
\end{center}
\end{figure}

The nature of the quasiparticle excitation is clarified a little
at the other parameter set studied, namely $\kappa=0.3620027$ (with all other
parameters unchanged). In this case our 
results show no evidence for any well-defined spin-${1\over2}$
state in either normal or anomalous channels; as a result of confinement
the excitation spectrum of the model seems to be saturated by the tightly
bound spin-0 states of Fig.~\ref{fig:bosons}.
Fig.~\ref{fig:wiggles}
shows a close up of $A(t)$ for various $\mu$ values, showing the presence of an
oscillatory component whose amplitude initially grows with $\mu$ 
(the $\tilde\mu=0.3$
points overlay those from $\tilde\mu=0.2$), but whose wavelength is roughly
$\mu$-independent. The origin of the oscillation could possibly be associated
with the non-unitarity of the model discussed in Sec.~\ref{sec:4D}, but is most
likely a manifestation of independent spin-${1\over2}$ excitations being
ill-defined due to confinement. 
We have fitted the $\kappa=0.3620027$ data to the forms
\begin{eqnarray}
N(t)=Pe^{-E_Nt}\cos(\Gamma_Nt+\phi)+Qe^{-E_N(L_t-t)}\cos(\Gamma_N(L_t-t)+\phi)
,\label{eq:fitNosc}\\
A(t)=R[e^{-E_At}\cos(\Gamma_At+\phi)
-e^{-E_A(L_t-t)}\cos(\Gamma_A(L_t-t)+\phi)],\label{eq:fitAosc}
\end{eqnarray}
where we interpret $E$ as the energy and $\Gamma$ as the width of a
quasiparticle excitation.
The results are plotted in Fig.~\ref{fig:mgamma}.
Most of the results are taken in Coulomb gauge, but at $\tilde\mu=0.3$ 
data from unitary gauge is
also available. The results' most striking feature is their independence
of $\mu$, with $\Gamma$ of the same order of magnitude as $E$. 
An interesting systematic
effect is that $E_N>E_A$ while $\Gamma_N<\Gamma_A$, which has motivated us in
Fig.~\ref{fig:modmgamma} to plot $\sqrt{E^2+\Gamma^2}$ vs. $\mu$. The figure has
the same vertical scale as the previous two, and it is clear that the disparity
between normal and anomalous channels is significantly reduced. Inspection of
the $\tilde\mu=0.3$ data also shows that the gauge dependence of this result is
O(20\%) at worst. Numerically, $\sqrt{E^2+\Gamma^2}>m_\pi$, indicating strong 
quark -- anti-quark binding, due to the persistence of confinement at this value
of $\kappa$.
Therefore we can interpret the effect of confinement 
as rotating
the quasiparticle pole into the complex plane, the rotation angle being larger
in the anomalous channel than in the normal one. It would be interesting if this
feature could be reproduced by analytic methods such as self-consistent solution
of Schwinger-Dyson equations.

\section{Conclusion}
\label{sec:conclusions}

Our attempt to alter the nature of the gluon background by changing the
parameters of the 3$d$ DR gauge-Higgs model has been a partial success, in that
in going from $\kappa=0.3620027$ to $\kappa=0.1$ 
the strength of the binding between quarks weakens significantly. 
The main evidence for this claim
comes from the
spectrum; at the smaller $\kappa$ studied quasiquarks appear to be well-defined
independent degrees of freedom, whose excitation energy is to fair precision 
half that 
of the gauge-invariant pion state. At $\kappa=0.3620027$ by
contrast, the quasiquark propagator exhibits a pole at complex $k$, and
the resulting estimates for both energy and width of the excitation exceed the
pion mass. However, in neither case is there evidence for significant departure
of $n_q$, $\langle\bar qq\rangle$ and $\langle qq\rangle$ from the behaviour 
predicted by $\chi$PT, so that even if quarks are important degrees of 
freedom at $\kappa=0.1$, there is no evidence for the formation of
a degenerate system signalled by SB scaling. 
Moreover, our attempts to measure the quasiquark
dispersion relation $E(k)$ have been unsuccessful; while there is evidence for 
the gauge-invariance of the minimum energy or ``gap'' at $\kappa=0.1$, no
$k$-dependence is observed. This is, of course, entirely consistent with 
the well-known fact that the only gauge invariant feature of a propagator is
the location of its poles. We conclude that identification of a Fermi surface,
supposing one existed, presents a technical challenge in any gauge theory 
simulation.

It is clear that our simplistic treatment, despite being manifestly gauge
invariant and inclusive of non-pointlike interactions,
 has missed some essential component
of the physics of high density. 
If we had 
first attempted the quenching programme using gauge group SU(3), the
departures from the theoretical expectation of SB scaling in the large-$\mu$
limit could have been ascribed either to unexpected non-perturbative effects, or
to some subtle cancellation due to the Sign Problem entirely missing
from the quenched approach. For SU(2), however, orthodox simulations 
of full QC$_2$D, in which the Sign Problem is absent, yield evidence for
deconfinement and SB scaling at large $\mu$~\cite{HKS,ADL}, which our approach
misses completely. 
We can of course speculate on which important features of
the gluon background at high quark density are absent; 
possible candidates are non-static modes, and modes with $\vert\vec
k\vert\approx2k_F\approx2\mu$~\cite{friedel}. Sadly though, it appears to remain
the case that despite its ``unreasonable effectiveness'' in virtually
all other aspects of lattice QCD, the quenched
approximation has nothing useful to tell us about the physics of high quark
density.

\section*{Acknowledgements}
We thank Owe Philipsen for generously giving us access to the gauge-Higgs
simulation code. The numerical work was performed on the 
PC cluster ``Majorana'' of the ``INFN~-~Gruppo Collegato di Cosenza'' 
at the ``Universit\`a della Calabria'' (Italy).

\end{document}